\documentclass[a4paper,11pt]{article}

\usepackage{aas_macros}
\usepackage{physics}
\usepackage[margin=2cm]{geometry}
\usepackage{graphicx}
\usepackage{subcaption}
\usepackage{amsmath,amssymb}
\usepackage{hyperref}
\usepackage{multirow}
\usepackage{rotating}
\usepackage[dvipsnames]{xcolor}

\graphicspath{{figures/}} 

\usepackage[dvipsnames]{xcolor}
\hypersetup{
colorlinks=true,
urlcolor=ForestGreen,
linkcolor=blue,
citecolor=magenta}

\title{Cosmology-informed Neural Networks to infer dark energy equation-of-state}
\author{Anshul Verma$^1$, Shashwat Sourav$^2$, Pavan K. Aluri$^1$, David F. Mota$^3$}
\date{\today}

\begin{document}

\maketitle

\centerline{$^1$Department of Physics, Indian Institute of Technology (BHU), Varanasi - 221005, India}
\centerline{$^2$Department of Physics, Washington University in St. Louis, MO 63130, USA}
\centerline{$^3$Institute of Theoretical Astrophysics, University of Oslo, Oslo, P.O.Box 1029 Blindern, N-0315, Norway}

\begin{abstract}

We present a framework that combines physics‐informed neural networks (PINNs) with Markov Chain Monte Carlo (MCMC) inference to constrain dynamical dark energy models using the Pantheon+ Type Ia supernova compilation. First, we train a physics-informed neural network to learn the solution of the Friedmann equation and accurately reproduce the matter density term $x_m(z)=\Omega_{m,0}(1+z)^3$ across a range of $\Omega_{m,0}$. For each of five two‐parameter equation-of-state (EoS) forms : Chevallier-Polarski-Linder (CPL), Barboza-Alcaniz (BA), Jassal-Bagla-Padmanabhan (JBP), Linear-$z$, and Logarithmic-$z$, we derive the analytic dark energy factor $x_{\rm de}(z)$, embed the trained surrogate within a GPU‐accelerated likelihood pipeline, and sample the posterior of $(h_0,\Omega_{m,0},w_0,w_a,M_0)$ using the \texttt{emcee} ensemble sampler with the full Pantheon+ covariance. All parameterizations remain consistent with a cosmological constant ($w_0=-1$, $w_a=0$) at the 95\% credible level, with the tightest bounds from the CPL form. While the surrogate does not reduce computation time for a single run in simple models, it becomes advantageous for repeated analyses of the same EoS or for models with expensive likelihood evaluations, and can be shared as a reusable with different datasets within the training range of SNe redshifts. This flexibility makes the approach a scalable tool for future cosmological inference, especially in regimes where conventional ODE‐based methods are computationally prohibitive.

\end{abstract}

\section{Introduction}

Artificial neural networks (ANNs \cite{Goodfellow-et-al-2016,yegnanarayana2009artificial,Uhrig_ANN_1995,krenker2011introduction}) have evolved from purely data‑driven pattern recognizers into flexible function approximators that can obey physical constraints exactly. Physics‑informed neural networks (PINNs) \cite{raissi2019physics} embody this idea by minimising the residuals of the governing differential equations, rather than comparing against pre‑computed labels, and thus act as continuous, differentiable surrogate models\footnote{A surrogate model is a fast, differentiable approximation that replaces repeated expensive numerical integrations.}. 
These features are highly advantageous in physics and cosmology, where solutions are needed across an entire domain and their derivatives feed into observable calculations or sensitivity analyses \cite{2025PhLB..86839690B}. Early cosmological applications reconstructed the Hubble expansion history directly from distance data \cite{2015JCAP...09..045V} and emulated Boltzmann codes for power‑spectrum inference \cite{2022MNRAS.512L..44S}. A key advance was the bundle‑solution scheme introduced in Ref.~\cite{2020arXiv200614372F}, in which a single network is trained over the entire prior hyper‑cube\footnote{The hyper‑cube is the multi‑dimensional box that bounds the allowed parameter ranges.}, so the costly ordinary‑differential‑equation (ODE) integration is performed only once.  
A subsequent work \cite{2023cinn} demonstrated that such bundle PINNs reproduce the background dynamics of several dark‑energy scenarios to sub‑percent accuracy, enabling their direct use inside Markov chain Monte Carlo (MCMC) samplers for parameter inference.

Our work extends that programme by (i) adding a dedicated validation layer, (ii) clarifying the mathematical structure of each equation-of-state (EoS) parameterization, and (iii) fusing data likelihoods and network outputs into a unified Bayesian engine.

Since the discovery of cosmic acceleration via observations of Type Ia supernovae (SNe Ia)~\cite{Perlmutter:1999,Riess:2021shoes,2022ppluscosmo}, Considerable effort has been directed towards constructing and testing theoretical models that could explain the late-time acceleration of the Universe. The standard cosmological model, known as the $\Lambda$CDM model, attributes this accelerated expansion to a cosmological constant, $\Lambda$, representing a constant vacuum energy density. Despite its remarkable success in explaining a wide array of cosmological observations, the $\Lambda$CDM model is not without its conceptual and observational challenge. From a theoretical point of view, the most pressing issue is the so-called cosmological constant problem, which highlights the enormous discrepancy between the observed value of $\Lambda$ and the theoretical expectations from quantum field theory~\cite{CarrollPressTurner1992,2000astro.ph..5265W,Carroll2001LRR,PeeblesRatra2003RvMP,2013arXiv1309.4133B,2014arXiv1407.2086M}. On the observational side, tensions have emerged when inferring the value of the Hubble constant `$H_0$' from different datasets. In particular, the SH0ES Collaboration's direct local measurements of $H_0$~\cite{Riess:2021shoes} differ significantly (at a level of $\sim 5\sigma$) from the value inferred from early Universe probes, specifically CMB measurements by the Planck Collaboration, under the assumption of $\Lambda$CDM \cite{Planck:2018cosmopar}. This discrepancy, known as the Hubble tension~\cite{Efstathiou:2020,freedman2021}, has sparked interest in exploring extensions or alternatives to the standard model that may reconcile these measurements (see for example~\cite{valentino:2021,2023Kamionkowski}).

Among the most promising areas to address the limitations of the $\Lambda$CDM model is the exploration of time-varying dark energy through phenomenological parameterizations of its equation-of-state. These so-called parametric dark energy models assume a redshift-dependent form for the equation-of-state parameter, $w(z)$, enabling one to probe possible deviations from a cosmological constant in a relatively model-independent way. Such parameterizations offer a tractable way to capture the dynamics of dark energy without committing to a specific underlying theory, and they are particularly well-suited to analyses involving low-redshift observables such as SNe Ia.

In this work, we focus on a suite of widely studied $w(z)$ parameterizations, each with its own motivation and distinctive redshift behaviour:

\begin{itemize}
    \item \textbf{Chevallier–Polarski–Linder (CPL)} \cite{2001IJMPD..10..213C,2003PhRvL..90i1301L}: The CPL parameterization is perhaps the most widely studied in literature. This two-parameter extension of the dark energy equation-of-state, given by $w(z) = w_0 + w_a \frac{z}{1+z}$ captures low-redshift evolution while remaining well-behaved at high redshifts. This parameterization offers several practical and theoretical advantages in modeling time-varying dark energy. First, it reduces the description of dark energy evolution to a manageable two-dimensional parameter space $(w_0, w_a)$, making it suitable for statistical inference. At low redshift, it recovers the commonly used linear approximation of $w(z)$, while at high redshift it remains bounded, avoiding divergences often present in simpler forms. Moreover, the CPL form has been shown to accurately reconstruct a wide class of scalar field models and their associated distance–redshift relations \cite{2004PhRvD..70h3007S,linden2008test,raveri2020reconstructing}. Its sensitivity to observational data makes it particularly effective in cosmological parameter estimation. Beyond technical convenience, this parameterization allows exploration of the expansion history deep into the pre-recombination era \cite{alam2021completed,lodha2025extended}, enabling constraints not only on dark energy but also on the broader dynamics of the Universe.

    \item \textbf{Barboza–Alcaniz (BA)} \cite{2008PhLB..666..415B}: The Barboza–Alcaniz (BA) parameterization was introduced as a phenomenological model for a time-varying dark energy equation-of-state that remains well-behaved across the full redshift range of cosmological evolution. Its functional form, $ w(z) = w_0 + w_a \frac{z(1+z)}{1+z^2}$ ensures that $w(z)$ remains bounded for all $z \in [-1, \infty)$, in contrast to several earlier models that diverge at high redshift or in the distant future. This property allows the BA model to be reliably applied from the last scattering surface ($z \sim 1100$) to the far future ($z = -1$), making it a useful tool for studying the full temporal behaviour of dark energy. The original analysis categorized different physical regimes such as quintessence ($w > -1$), phantom ($w < -1$), and models that cross the so-called phantom divide (i.e., $w=-1$) within the parameter space $(w_0, w_a)$. Through joint constraints using data from Type Ia supernovae, BAO, the CMB shift parameter (that is related to the location of first acoustic peak in CMB power spectrum), and $H(z)$ measurements, the model demonstrated observational compatibility with both quintessence and phantom-like behaviours, suggesting that a dynamic dark energy component is still viable. The BA model offers a flexible and observationally consistent framework for exploring late-time cosmological dynamics.

    \item \textbf{Jassal–Bagla–Padmanabhan (JBP)} \cite{2003PhRvD..67f3504B,2005MNRAS.356L..11J}: The JBP parameterization introduces a redshift-dependent equation-of-state for dark energy of the form: $w(z) = w_0 + w_a \frac{z}{(1 + z)^2}$ designed specifically to capture the dynamics of dark energy at intermediate redshifts ($z \sim 1$), where it most strongly influences the expansion history. One of the key features of this model is that it asymptotically approaches $w_0$ at both low ($z \to 0$) and high ($z \to \infty$) redshifts, unlike the CPL form which evolves unbounded at early times. This makes the JBP form particularly well-suited for probing deviations from $\Lambda$CDM that may occur only temporarily during cosmic evolution, while still maintaining consistency with early and late-time observational constraints. Motivated by the lack of strong evidence for dark energy evolution from early-universe probes like the CMB, the JBP model confines most of the deviation in $w(z)$ to redshifts around the onset of acceleration, minimizing its impact at recombination and ensuring compatibility with early-Universe physics. In their original analysis, Jassal et al.~\cite{2003PhRvD..67f3504B} demonstrated that the model fits SNe Ia data comparably to $\Lambda$CDM while allowing for time-varying behaviour in a narrow redshift window. The JBP parameterization serves as a phenomenological probe of transient dark energy evolution, offering a complementary perspective to other parameterizations like CPL and BA that allow broader redshift evolution.  

    \item \textbf{Linear parameterization} \cite{2001PhRvD..64l3527H,2002PhRvD..65j3512W}: The linear parameterization of the dark energy equation-of-state is among the earliest phenomenological models proposed to capture deviations from a cosmological constant. In its standard form, the EoS is written as: $ w(z) = w_0 + w_a z , w_a \equiv \left. \frac{dw}{dz} \right|_{z=0}$ which introduces a linear dependence on redshift. This simple structure enables straightforward implementation and interpretation, making it a useful testbed for probing dark energy dynamics over small to intermediate redshift ranges. However, the model diverges at large $z$, which limits its applicability in analyses involving early-universe data such as the CMB. In a more recent work \cite{2018PhRvD..97l3525D}, a linear form within the framework of unified dark sector models was employed. Their analysis showed that even this elementary parameterization can provide viable fits to observational data when treated carefully within a limited redshift domain. The primary motivation is to retain model simplicity while exploring whether a unified description of dark matter and dark energy can mimic standard $\Lambda$CDM-like behaviour and capture possible low-redshift deviations. Despite its limitations at high redshift, the linear parameterization remains instructive for late-time studies where data is primarily drawn from Type Ia supernovae and cosmic chronometers.

    \item \textbf{Logarithmic parameterization} \cite{Feng_2011,2011PhLB..699..233M,2017JCAP...06..012T}: Given by $w(z) = w_0 + w_a \ln(1+z)$, where $w_a \equiv \left. \frac{dw}{dz} \right|_{z=0}$, this model grows logarithmically with redshift and provides a slower and milder evolution of the equation-of-state compared to linear models. It is motivated by its appearance in some scalar field scenarios and offers an alternative to polynomial expansions when examining small deviations from $w = -1$. The logarithmic parameterization of the dark energy equation-of-state offers a smooth and slowly evolving alternative to more commonly used linear or CPL forms. In contrast to the linear parameterization, which diverges at large redshift, this model also has poor behaviour at $z>>1$ and becomes infinite . This dark energy EoS parameterization can only describe the behaviour of dark energy when $z$ is not very large. Physically, this parameterization allows for a slowly varying dark energy component that could deviate from $\Lambda$CDM in a gradual, cumulative manner over cosmic time, given there is no interaction between dark energy and other component of the universe. It is especially well-suited for probing subtle deviations from $w = -1$ in SNe Ia and Hubble expansion datasets. By introducing a weak redshift dependence in $w(z)$, the logarithmic model helps test whether a slowly evolving dark energy can better fit observational data compared to static or strongly dynamical alternatives.

\end{itemize}

These parameterizations collectively span a range of behaviours, including monotonic and non-monotonic evolutions, bounded and unbounded forms, and various asymptotic limits at high and low redshifts. Rather than solving the background equations numerically for each model, we adopt a model-agnostic approach by first constructing a neural network solution to the Hubble expansion rate $H(z)$ using Pantheon+ SNe Ia data calibrated with SH0ES, assuming a fiducial $\Lambda$CDM cosmology. This ANN-based $H(z)$ function serves as an effective observational proxy, allowing us to test the consistency of various $w(z)$ parameterizations without re-solving the cosmological equations for each case. Subsequently, we perform Bayesian inference using Markov Chain Monte Carlo (MCMC) techniques to constrain the parameters of each $w(z)$ model, and assess their agreement with the ANN-reconstructed expansion history. In doing so, we aim to determine whether any of these models exhibit statistically significant deviations from the $\Lambda$CDM baseline, and to characterize the extent to which late-time dynamics of dark energy are supported by current observations.

Here, our primary goal is to demonstrate a practical and efficient method for cosmological model testing based on an ANN-based bundle solution of the expansion history, trained under the $\Lambda$CDM and five different $\omega_a \omega_b$CDM assumptions. This ANN reconstruction acts as a numerical surrogate for the Hubble rate $H(z)$, allowing for rapid and repeated evaluations during likelihood-based parameter inference. We then apply this approach to constrain several time-dependent dark energy parameterizations using the Pantheon+ SNe Ia dataset with SH0ES calibration. This framework avoids repeated numerical integration of the Friedmann equation for each parameterization, significantly reducing the computational cost of exploring a broad class of models. Rather than uncovering model-agnostic signals of new physics, our focus is on validating the utility and consistency of this approach in the context of standard parametric dark energy modeling.

The rest of the paper is organized as follows. In Section~\ref{sec:eos} we present the explicit analytic forms of the five dynamical dark energy equation-of-state parameterizations considered in this work, namely CPL, BA, JBP, Linear-$z$, and Logarithmic-$z$. Section~\ref{sec:ann_method} outlines our artificial neural network methodology, including the physics-informed training procedure and the overall end-to-end likelihood pipeline. In Section~\ref{sec:ann_surrogate_chronology} we describe the surrogate construction strategy in detail, covering analytic boundary enforcement, network architecture, residual loss formulation, and training procedure. The observational dataset and likelihood analysis setup are described in Section~\ref{sec:SNIa-data-likelihood}. Our main findings are presented in Section~\ref{sec:results}, where we report posterior constraints for all models, quantify the accuracy of the ANN surrogates, and assess the computational efficiency of the CINN approach. We summarize our conclusions in Section~\ref{conclusion} and discuss potential future applications. Additional discussion of computational cost scaling is provided in the Appendix.

\section{Dark energy EoS parameterizations : Explicit analytic forms}
\label{sec:eos}

A commonly used approach to explore deviations from the standard cosmological constant scenario is to generalize the dark energy equation-of-state (EoS) to a redshift-dependent function, defined as \( w(z) = p_{\rm de}/\rho_{\rm de} \), where \( p_{\rm de} \) and \( \rho_{\rm de} \) are the pressure and energy density of dark energy, respectively. This allows for a time-varying behaviour of dark energy, which may offer a better fit to current and future observations.

We assume a homogeneous and isotropic Universe described by the spatially flat Friedmann-Robertson-Walker (FRW) metric,
\begin{equation}
\label{eq:frw_metric}
    ds^2 = dt^2 - a(t)^2 \left( dx^2 + dy^2 + dz^2 \right),
\end{equation}
where \( a(t) \) is the scale factor and spatial flatness is assumed in accordance with current cosmological observations. Using this metric, the Einstein field equations reduce to the Friedmann equations. The first Friedmann equation governs the expansion rate and is given by
\begin{equation}
    H^2(t) = \left( \frac{\dot{a}}{a} \right)^2 = \frac{8\pi G}{3} \rho_{\text{tot}},
\end{equation}
where \( H(t) \) is the Hubble parameter and \( \rho_{\text{tot}} \) is the total energy density of the Universe.

In the late Universe, we consider a two-component system composed of pressureless matter (with energy density \( \rho_m \propto a^{-3} \)) and a dark energy component characterized by a redshift-dependent equation-of-state \( w(z) \). The behaviour of dark energy is governed by the conservation of energy-momentum, which yields the continuity equation:

\begin{equation}
    \frac{d\rho_{\rm de}}{dz} = \frac{3\,\rho_{\rm de}\,[1+w(z)]}{1+z},
    \label{eq:de_conservation}
\end{equation}
which has the formal solution
\begin{equation}
    \rho_{\rm de}(z) = \rho_{\rm de,0}
    \exp\!\left[3\int_{0}^{z}\!\frac{1+w(z')}{1+z'}\,dz'\right],
    \label{eq:rho_general}
\end{equation}
where \( \rho_{\rm de,0} \) is the present-day dark energy density.

For convenience, we define the dimensionless function
\(x_{\rm de}(z) \equiv \rho_{\rm de}(z)/\rho_{\rm de,0} \),
which encapsulates the redshift evolution of the dark energy density. In terms of this auxiliary function, the Hubble parameter becomes
\begin{equation}
    H^{2}(z) = H_0^{2}\left[
    \Omega_{m,0}(1+z)^{3}
    + (1-\Omega_{m,0})\,x_{\rm de}(z)
    \right],
    \label{eq:friedmann_1}
\end{equation}
where \( \Omega_{m,0} \) is the present-day matter density parameter, and spatial flatness is assumed.

In this work, we consider five widely studied analytic forms of \( w(z) \), each designed to generalize \(\Lambda\)CDM in a specific way while introducing one or two free parameters beyond \( w = -1 \). Each model leads to an exact expression for \( x_{\rm de}(z) \) when inserted into Eq.~(\ref{eq:rho_general}). The equation-of-state corresponding to different models used in this work are given as:

\paragraph*{(i) Chevallier–Polarski–Linder (CPL):}
This parameterization expands \( w(z) \) linearly in scale factor \( a = 1/(1+z) \), yielding
\[
w(z) = w_0 + w_a \frac{z}{1+z}.
\]
It ensures smooth behaviour at both low and high redshift, approaching \( w_0 + w_a \) as \( z \to \infty \). The resulting dark energy density evolution is
\begin{equation}
    x_{\rm de}^{\rm CPL}(z)=(1+z)^{3(1+w_0+w_a)}
    \exp\!\left[-\frac{3w_a z}{1+z}\right].
\label{eq:xde-cpl}
\end{equation}

\paragraph*{(ii) Barboza–Alcaniz (BA):}
This form is symmetric about \( z = 1 \) and is given by
\[
w(z) = w_0 + w_a \frac{z(1+z)}{1 + z^2}.
\]
It is bounded and well-behaved at all redshifts, approaching \( w_0 + w_a \) at both high and low \( z \). The corresponding evolution is
\begin{equation}
    x_{\rm de}^{\rm BA}(z) =
    (1+z)^{3(1+w_0)}
    \left(1+z^{2}\right)^{\tfrac{3}{2}w_a}.
\label{eq:xde-ba}
\end{equation}

\paragraph*{(iii) Jassal–Bagla–Padmanabhan (JBP):}
The JBP form,
\[
w(z) = w_0 + w_a \frac{z}{(1+z)^2},
\]
is designed to allow deviation from \( w = w_0 \) primarily at intermediate redshifts, with \( w(z) \to w_0 \) at both \( z = 0 \) and \( z \to \infty \). The solution is
\begin{equation}
    x_{\rm de}^{\rm JBP}(z) =
    (1+z)^{3(1+w_0)}
    \exp\!\left[\frac{3w_a z^{2}}{2(1+z)^{2}}\right].
\label{eq:xde-jbp}
\end{equation}

\paragraph*{(iv) Linear-$z$ model:}
A simple Taylor expansion in redshift,
\[
w(z) = w_0 + w_a z,
\]
provides maximal flexibility at low redshifts, where most observational data lie. However, it diverges as \( z \to \infty \), which limits its applicability at early times. The evolution reads
\begin{equation}
    x_{\rm de}^{\rm Lin}(z) =
    (1+z)^{3(1+w_0 - \tfrac{w_a}{2})}
    \exp\!\left[\tfrac{3}{2}w_a z\right].
\label{eq:xde-lin-z}
\end{equation}

\paragraph*{(v) Logarithmic-$z$ model:}
This form introduces a mild evolution in \( w(z) \),
\[
w(z) = w_0 + w_a \ln(1+z),
\]
which remains perturbative across cosmic time. Its density evolution is
\begin{equation}
    x_{\rm de}^{\rm Log}(z) =
    (1+z)^{3(1+w_0)}
    \exp\!\left[\frac{3w_a}{2}\ln^{2}(1+z)\right].
\label{eq:xde-log-z}
\end{equation}

Each of these parameterizations reduces to \(\Lambda\)CDM for \( w_0 = -1 \) and \( w_a = 0 \), thereby offering a consistent baseline for comparison across the models. This allows us to test whether current observational data support dynamical dark energy or remain consistent with a constant vacuum energy. In what follows, we describe a methodology for evaluating these models efficiently: instead of solving the background equations for each parameterization separately, we construct a single, data-driven approximation to the expansion history using artificial neural networks (ANNs). This ANN-based framework enables rapid inference across the entire parameter space while maintaining high accuracy, and forms the core of our model comparison strategy.

\section{Artificial Neural Network Methodology}
\label{sec:ann_method}

In this section, we describe the core methodology behind our work: the use of an artificial neural network (ANN) as a surrogate model for solving the cosmological background equations. Instead of solving the Friedmann equation numerically for each parameterization of the dark energy equation-of-state (EoS), we train a Physics-Informed Neural Network (PINN) to learn the solution to the governing differential equation for dark energy evolution. Once trained, the ANN provides a fast and differentiable approximation of the expansion history for any given set of model parameters. This surrogate replaces repeated numerical integrations in the Bayesian inference pipeline, significantly improving computational efficiency. The framework applies to five dark energy EoS models detailed in Sec.~\ref{sec:eos}: $\Lambda$CDM, Barboza–Alcaniz (BA), JBP, Linear, and Logarithmic parameterizations. For each model, we define the EoS \( w(z; \theta) \), where \( \theta = \{w_0, w_a\} \) are model-specific parameters. The goal is to learn the redshift evolution of the normalized dark energy density,
\begin{equation}
x_{\text{de}}(z) \equiv \frac{\rho_{\text{de}}(z)}{\rho_{\text{de},0}},
\end{equation}
without directly solving the Friedmann equation for each parameter set during inference.

We begin by rewriting the energy conservation equation (Eq.~\ref{eq:de_conservation}) in terms of \( x_{\text{de}}(z) \). This yields the first-order linear ordinary differential equation (ODE):
\begin{equation}
    \frac{dx_{\text{de}}}{dz} - \frac{3[1+w(z;\theta)]}{1+z}\,x_{\text{de}} = 0,
    \label{eq:master_ode}
\end{equation}
with the initial condition \( x_{\text{de}}(0) = 1 \). This ODE encapsulates the full cosmological evolution of the dark energy density for a given EoS model.

\paragraph*{Surrogate Learning via PINNs:}
To learn the solution of Eq.~(\ref{eq:master_ode}), we train a neural network \( x_{\text{de}}(z; \theta) \) that takes redshift \( z \) and EoS parameters \( \theta \) as inputs and outputs the predicted value of \( x_{\text{de}} \). The network is trained by minimizing the residual of the ODE over a range of redshifts and parameter values, using the physics-informed loss:
\begin{equation}
    L_{PINN} = \left\langle \left( \frac{d}{dz} x_{\text{de}}(z;\theta) - \frac{3[1+w(z;\theta)]}{1+z} x_{\text{de}}(z;\theta) \right)^2 \right\rangle_{z,\theta}.
\end{equation}
Here, \( \langle \cdot \rangle \) denotes averaging over the training dataset, which consists of randomly sampled points in the \( z \)–\( \theta \) space. The derivative \( d x_{\text{de}}/dz \) is computed via automatic differentiation \cite{2024arXiv240514099C}.

This procedure yields what we refer to as a \textit{bundle solution}: a continuous map from model parameters \( \theta \) to the function \( x_{\text{de}}(z) \), valid throughout the training domain. Once trained, the surrogate model can be evaluated within half an hour and can replace numerical solvers in downstream analysis.

\begin{figure}
    \centering
    \includegraphics[width=0.95\textwidth]{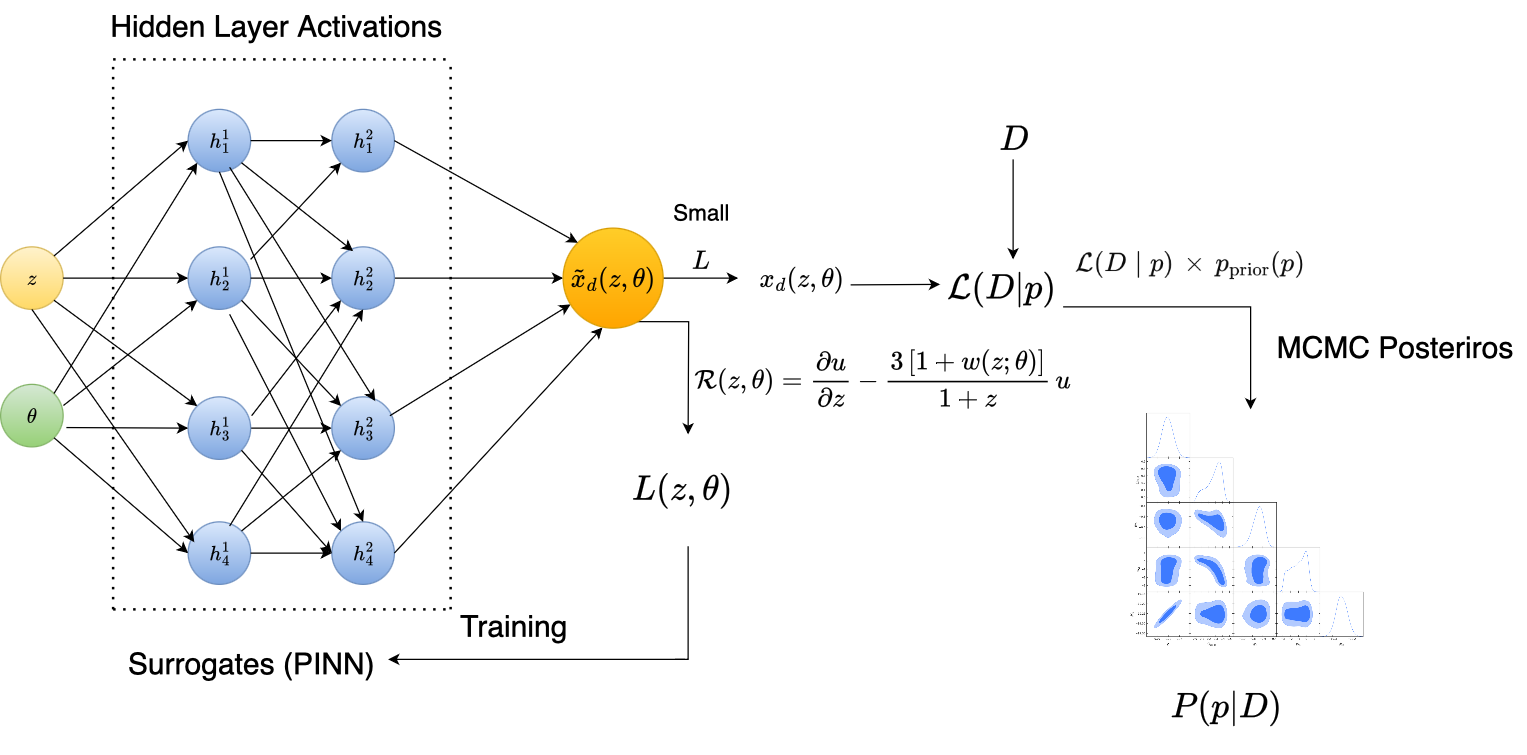}
    \caption{PINN–MCMC pipeline. \textbf{Left:} The PINN takes redshift \( z \) and EoS parameters \( \theta \) as inputs, passes them through two hidden layers to output \( u(z, \theta) \), from which \( x_{\rm de}(z, \theta) = \exp(u) \) is reconstructed. The residual
    \(
    \mathcal{R}(z, \theta) = \partial_z u - 3[1 + w(z; \theta)]/(1 + z)
    \)
    is minimized. \textbf{Right:} The trained surrogate feeds into the Friedmann equation to compute \( H(z) \), which is used in the cosmological likelihood. MCMC is then performed to infer the posteriors.}
    \label{fig:ann_mcmc_pipeline}
\end{figure}

\paragraph*{End-to-End Pipeline:}
The left panel of Fig.~[\ref{fig:ann_mcmc_pipeline}] illustrates the training process: redshift \( z \) and EoS parameters \( \theta \) are input into the ANN, which returns \( x_{\text{de}}(z;\theta) \). The output is constrained to satisfy the ODE in Eq.~(\ref{eq:master_ode}) via the residual loss (described later).

After training, the ANN acts as a fast cosmological engine inside the likelihood evaluation. As shown in the right panel of Fig.~[\ref{fig:ann_mcmc_pipeline}], we feed the ANN output into the Hubble function \( H(z;\theta) \) using the modified Friedmann equation (Eq.~\ref{eq:friedmann_1}), compute distance moduli, and compare to observational data via a standard likelihood function. We then perform Bayesian inference using Markov Chain Monte Carlo (MCMC) to obtain posterior distributions on \( \theta \), with the trained ANN called repeatedly at each MCMC step.

This surrogate-based approach enables rapid evaluation of cosmological models across parameter space and avoids the need to re-integrate differential equations at every likelihood call. It is particularly well-suited for families of analytic \( w(z) \) models with tractable redshift dependence but expensive numerical evaluation.

\subsection{Surrogate Construction Strategy}
\label{sec:ann_surrogate_chronology}

For each of the five EoS models ($\Lambda$CDM, BA, JBP, Linear, Log), we construct a surrogate network to approximate \( x_{\rm de}(z) \) as a function of redshift and EoS parameters (e.g., \( w_0, w_a \)). The contribution from matter is known analytically as \( x_m(z) = \Omega_{m,0}(1 + z)^3 \), so the total expansion rate is reconstructed as:
\begin{equation}
E^2(z) = x_m(z) + (1 - \Omega_{m,0}) x_{\rm de}(z).
\end{equation}

For clarity, we distinguish the three components that recur in our validation tests:
\begin{itemize}
  \item \textbf{Matter term} \(x^{\text{ANN}}_{m}(z,\Omega_{m,0})\equiv\Omega_{m,0}(1+z)^3\).  We reproduce this analytic scaling with a tiny two-layer feed-forward network ($\mathcal{O}(10^2)$ weights) so that \emph{all} background quantities, matter and dark energy, live in a single differentiable PyTorch module.  The network is trained once against the closed-form target, reaches machine-precision (\(<10^{-6}\) MSE) in $\sim\!10^2$ Adam iterations, and its weights are then frozen.  Figure~\ref{fig:xm_error} therefore measures the residual of this auxiliary net, not a new free function.
  \item \textbf{Dark-energy term} \(x^{\text{PINN}}_{de}(z;\theta)\).  This is the bundle Physics-Informed Neural Network that solves \ref{eq:u_ode} for every EoS parameter pair \(\theta=(w_0,w_a)\) in one shot.
  \item \textbf{Dimensionless Hubble rate} \(E(z)=\sqrt{x^{\text{ANN}}_{m} + (1-\Omega_{m,0})x^{\text{PINN}}_{de}}\) and the distance modulus \(\mu(z)=5\log_{10}\!\bigl[(1+z)\chi(z)/\text{Mpc}\bigr]+25\), where \(\chi(z)=\int_0^{z}\!E^{-1}(z')\,dz'\).
\end{itemize}
Throughout Sec.~\ref{sec:accuracy_results} we validate these three layers in turn: (i) the stand-alone matter net (Fig.~\ref{fig:xm_error});   (ii) its propagation into \(E(z)\) (Fig.~\ref{fig:percent_error}) and (iii) the final impact on the observable \(\mu(z)\) (Fig.~\ref{fig:dm_models}). Because the matter surrogate is already accurate to \(\lesssim10^{-5}\), all reported cosmological-scale errors are dominated by the dark-energy PINN and remain orders of magnitude below current observational uncertainties.
The surrogate construction involves four key steps:

\paragraph*{(i) Analytic Scaffolding (Boundary Condition Enforcement):}
To ensure the boundary condition \( x_{\rm de}(0) = 1 \) is satisfied, we reparametrize the solution using an auxiliary function \( u(z) \) such that
\begin{equation}
    x_{\rm de}(z) = \exp[u(z)].
    \label{eq:xde_u_transform}
\end{equation}
This guarantees \( u(0) = 0 \) implies \( x_{\rm de}(0) = 1 \). Substituting into Eq.~(\ref{eq:master_ode}), we obtain a simpler ODE:
\begin{equation}
    \frac{du}{dz} = \frac{3[1 + w(z)]}{1 + z}.
    \label{eq:u_ode}
\end{equation}
This transformation shifts the learning target from the exponential behaviour of \( x_{\rm de} \) (see Eq.~\ref{eq:xde-cpl} - \ref{eq:xde-log-z}) to the smoother \( u(z) \), improving stability during training.

\paragraph*{(ii) Network Architecture:}
We define a neural network \( u_{\boldsymbol{\phi}}(z, \theta) \) that receives redshift \( z \) and EoS parameters (e.g., \( \{w_0, w_a\} \)) as input:
\begin{equation}
    u_{\boldsymbol{\phi}}: (z, \theta) \rightarrow \mathbb{R},
\end{equation}
where \( \boldsymbol{\phi} \) are the trainable weights and biases. The use of EoS parameters as explicit inputs enables the network to learn a \emph{bundle solution} across the parameter space. For $\Lambda$CDM, where \( w(z) = -1 \), the solution is trivial: \( u(z) = 0 \) and \( x_{\rm de}(z) = 1 \), which we handle analytically.

\paragraph*{(iii) Residual Loss and Physics-Informed Training:}
The network is trained by minimizing the residual of Eq.~(\ref{eq:u_ode}) across a set of collocation points \( (z_j, \theta_j) \). At each point, the residual is:
\begin{equation}
    \mathcal{R}_j(\boldsymbol{\phi}) = \frac{\partial u_{\boldsymbol{\phi}}}{\partial z}(z_j, \theta_j) - \frac{3[1 + w(z_j; \theta_j)]}{1 + z_j}.
    \label{eq:residual}
\end{equation}
As mentioned before, the derivative is computed via automatic differentiation. The total loss function is the mean squared residual over all training points, optionally including a boundary condition penalty which is simply the L2 regularization term \footnote{The additional \(L_2\) term penalises any deviation from the Dirichlet condition \(u(0)=0\), since the interior residual by itself cannot enforce this boundary. In practice we treat \(\lambda_{\text{BC}}\) as a tunable hyper‑parameter, adjusting it until the boundary error is comparable to the interior residual.}  \cite{2012arXiv1205.2653C}:
\begin{equation}
    \mathcal{L}(\boldsymbol{\phi}) = \frac{1}{N} \sum_{j=1}^N \mathcal{R}_j^2 (\boldsymbol{\phi}) + \lambda_{\text{BC}} \sum_k u_{\boldsymbol{\phi}}(0, \theta_k)^2.
    \label{eq:loss_pinn}
\end{equation}

\paragraph*{(iv) Training Procedure:}
We use a combination of gradient-based optimizers (typically Adam followed by L-BFGS) to minimize the loss. Training continues until the residual across the domain falls below a predefined tolerance (e.g., \( 10^{-6} \)). The loss is seen to reduce as we see residual gradually decrease below the tolrerance limit. This yields a surrogate capable of generating \( x_{\rm de}(z; w_0, w_a) \) for any combination of parameters in the training range.

Each of the five EoS models described in Sec.~\ref{sec:eos} defines a different functional form of \( w(z) \), but they all share the same master ODE structure. As a result, each EoS model admits its own dedicated PINN surrogate.

\section{Data set and likelihood analysis}
\label{sec:SNIa-data-likelihood}

For our analysis, we use the Pantheon+ Type Ia Supernovae (SNe Ia) compilation \cite{scolnic2022pantheon+}, which is currently the most comprehensive and homogeneous dataset of thermonuclear supernova distances. It supersedes the original Pantheon sample by increasing the number of events, improving photometric calibration, and incorporating updated corrections for peculiar velocities and host-galaxy biases.

The Pantheon+ compilation consists of 1701 light curves corresponding to 1543 unique spectroscopically confirmed SNe\,Ia, drawn from 18 independent surveys and spanning a redshift range of \(0.001 < z < 2.26\). The dataset includes duplicated observations of some SNe by multiple surveys, as well as ``SN siblings'' (multiple SNe discovered in the same host galaxy). Only well-measured and unambiguously classified Type Ia events are included, based on survey-specific quality criteria, such as the multi-tiered filtering in PS1~\cite{2022ApJ...938..113S} and comparable cuts in other datasets.
Light-curve fitting was performed using the SALT2 model
with photometric calibration anchored to the HST Calspec system \cite{coughlin2018testing,2018A&A...619A.180M}. The resulting observed distance moduli \( \mu_{\mathrm{obs},i} \) carry both statistical and systematic uncertainties.

The full SNe covariance matrix entering the likelihood is the sum of statistical covariances and systematic errors and is given as,
\begin{equation}
    \label{eq:covariance_matrix}
    \mathbf{C} = \mathbf{D}_{\sigma}^2 + \mathbf{C}_{\rm sys},
    \qquad (\mathbf{D}_{\sigma})_{ij} \equiv \sigma_{\mu,i}\,\delta_{ij},
\end{equation}
The Gaussian log–likelihood reads
\begin{equation}
\ln\mathcal L \;=\;
-\tfrac12\,
\bigl(\boldsymbol{\mu}_{\rm obs}-\boldsymbol{\mu}_{\rm th}\bigr)^T\,
\mathbf{C}^{-1}\,
\bigl(\boldsymbol{\mu}_{\rm obs}-\boldsymbol{\mu}_{\rm th}\bigr),
\label{eq:log-likelihood}
\end{equation}
where $\mathbf C$ is the \emph{fixed} $N\times N$ covariance matrix as given in Eq.~(\ref{eq:covariance_matrix}) ($N=1701$ for Pantheon+).

Here, \( \Delta \boldsymbol{\mu} = \mu^i_{\rm obs} - \mu^i_{\rm th} =  (\Delta \mu_1 \,\, \Delta \mu_2 \,\, \hdots \,\, \Delta \mu_N)^T \) is the residual vector between observed and model-predicted distance moduli, where `\( N \)' is the number of SNe\,Ia in the sample. For each supernova \( i \), the residual \( \Delta \mu_i \) is defined conditionally depending on whether the supernova is hosted in a Cepheid-calibrated galaxy or not as
\begin{equation}
    \Delta \mu_i =
    \left\{
    \begin{array}{ll}
    m_{B, \rm corr}^i - M_0 - \mu_i^{\rm Cepheid}, & \text{if } i \in \text{Cepheid host (calibrator)}, \\
    m_{B, \rm corr}^i - M_0 - \mu_i^{\rm th}(\boldsymbol{\phi}, z_i^{\ast}), & \text{otherwise},
    \end{array}
    \right.
    \label{eq:mu-residuals}
\end{equation}
where \( m_{B,\rm corr}^i \) 
is the SALT2-corrected apparent $B$-band magnitude for SNIa lightcurve stretch, colour, host galaxy mass bias, etc., and \( M_0 \) is the absolute magnitude of SNe\,Ia (treated as a nuisance parameter). In this work, we directly use the corrected magnitude values provided by the Pantheon+ collaboration under the column \verb+m_b_corr+ for each supernova for the data file provided. 
While the Pantheon+ dataset also provides uncorrected $B$-band magnitudes and SALT2 parameters (via the \verb+mB+, \verb+x1+, and \verb+c+, etc., columns), here we work with the pre-corrected \( m_{B, \rm corr} \) values marginalized over the corresponding nuisance parameters (SNIa lightcurve fit parameters).

For SNe in Cepheid-calibrated host galaxies (indicated by the binary column \verb+IS_CALIBRATOR+ = 1), the Cepheid-derived distance moduli \( \mu_i^{\rm Cepheid} \) are taken from the \verb+CEPH_DIST+ column. For all other SNe, the theoretical distance modulus \( \mu_i^{\rm th} \) is computed using the model-predicted Hubble function \( H(z; \boldsymbol{\phi}) \) dependent on the cosmological parameters \( \boldsymbol{\phi} \).

\begin{figure}
    \centering
    \includegraphics[height=4.4cm]{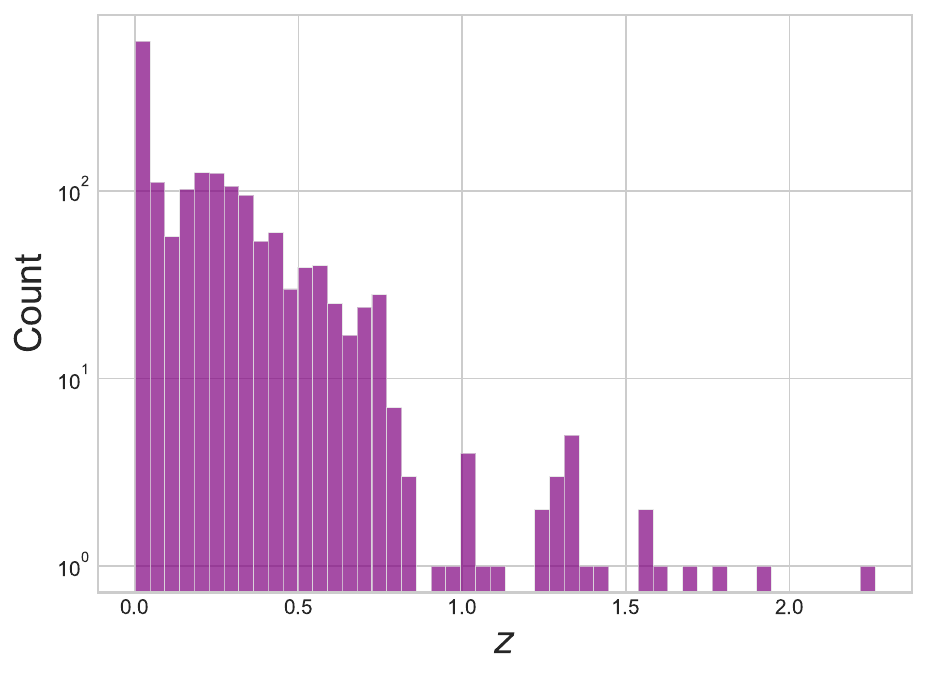}
    ~
    \includegraphics[height=4.6cm]{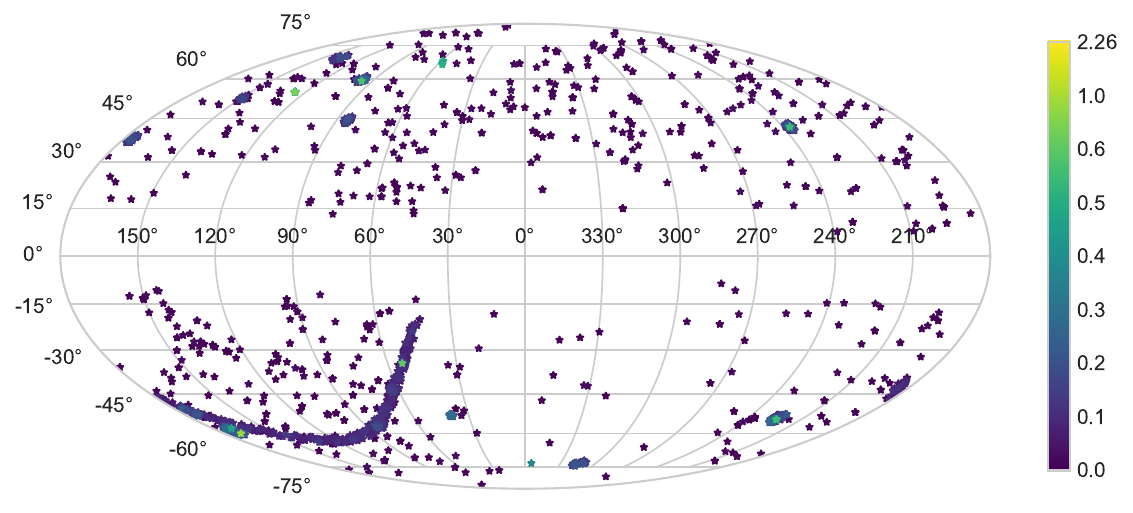}
    \caption{\emph{Left:} Redshift distribution of SNe\,Ia in the Pantheon+ sample, binned in 100 intervals. \emph{Right:} Sky distribution of the same sample, with colour bar denoting redshift.}
    \label{fig:pantheon+_dist}
\end{figure}

Although the Pantheon+ dataset does not directly allows us to constrain \( H_0 \) (due to its degeneracy with $M_0$), we note, however, that the full Pantheon+SH0ES framework~\cite{2022ppluscosmo} enables cosmological inference from a three-rung distance ladder: Cepheid-calibrated anchors, SN-host cross-calibration, and high-redshift Hubble-flow SNe. In this analysis, we adopt the same Pantheon+SH0ES-calibrated distance moduli for our surrogate-based likelihood evaluation.

We constrain the five-dimensional parameter vector $ \boldsymbol{\Theta} = \bigl(h_0, \Omega_{m,0}, w_0, w_a, M_0\bigr)$ using the observed distance moduli from the Pantheon+ SNe\,Ia dataset. The likelihood function given by Eq.~(\ref{eq:log-likelihood}) is 
evaluated by computing the residuals \( \Delta \mu_i \) for each supernova, as defined in Eq.~(\ref{eq:mu-residuals}). For a given dark energy EoS model (e.g., CPL, BA, etc.), the trained PINN provides the normalized dark energy density \( x_{\rm de}(z) \) as a function of redshift and EoS parameters : $w_0$ and $w_a$. The dimensionless Hubble parameter is then evaluated as
\[
E(z; \boldsymbol{\phi}) = \sqrt{\Omega_{m,0}(1+z)^3 + (1 - \Omega_{m,0}) x_{\rm de}(z)}.
\]
The comoving distance is computed by numerically evaluating the integral
\begin{equation}
\chi(z; \boldsymbol{\phi}) = \int_0^z \frac{dz'}{E(z'; \boldsymbol{\phi})},
\end{equation}
from which the theoretical distance modulus follow as
\begin{equation}
\mu_{\rm th}(z; \boldsymbol{\phi}) = 5 \log_{10} \left[ \frac{(1+z)\chi(z; \boldsymbol{\phi})}{h_0} \right] + 42.3841.
\end{equation}

For supernovae in Cepheid-calibrated host galaxies, the model-predicted modulus \( \mu_{\rm th} \) is replaced by the independent Cepheid-based host galaxy distance (per Eq.~\ref{eq:mu-residuals}). The full residual vector \( \Delta \boldsymbol{\mu} \) is then evaluated and used in the log-likelihood function as given in the Eq.~(\ref{eq:log-likelihood}).

We assume uniform priors over broad ranges: $ 0.55 \leq h_0 \leq 0.85$, $0.05 \leq \Omega_{m,0} \leq 0.60$, $-2 \leq w_0 \leq 0$, $-3 \leq w_a \leq 3$, and $-19.4 \leq M_0 \leq -19.1$, with prior ranges selected to lie entirely within the valid domain of the trained PINNs. Bayesian inference is performed using Markov Chain Monte Carlo (MCMC) with the \texttt{emcee} sampler \cite{2013PASP..125..306F}. We use an ensemble of walkers with appropriate burn-in and sampling steps to ensure convergence of each model. The resulting posterior distributions for all parameters are shown in Sec. \ref{sec:results}. We do not include any model-comparison statistics such as Bayes factors in this work, as our focus is on demonstrating the PINN-based surrogate framework and applying it consistently across various EoS models. A more detailed statistical comparison between models is deferred to future studies.

\section{Results}
\label{sec:results}

\subsection{Posterior Constraints}
\label{sec:posteriors}

\begin{figure}
\centering
\includegraphics[width=0.47\linewidth]{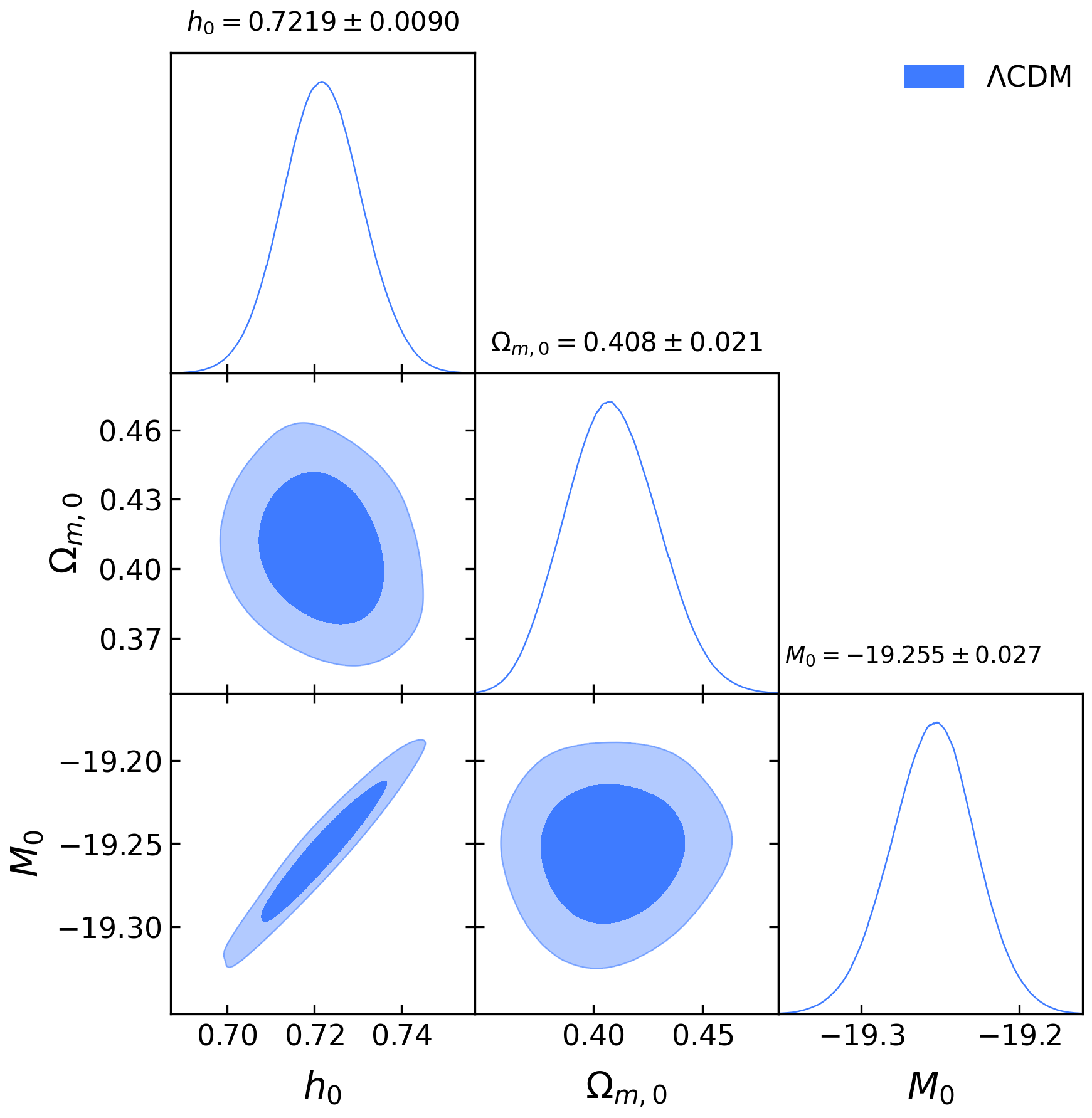}
~
\includegraphics[width=0.47\linewidth]{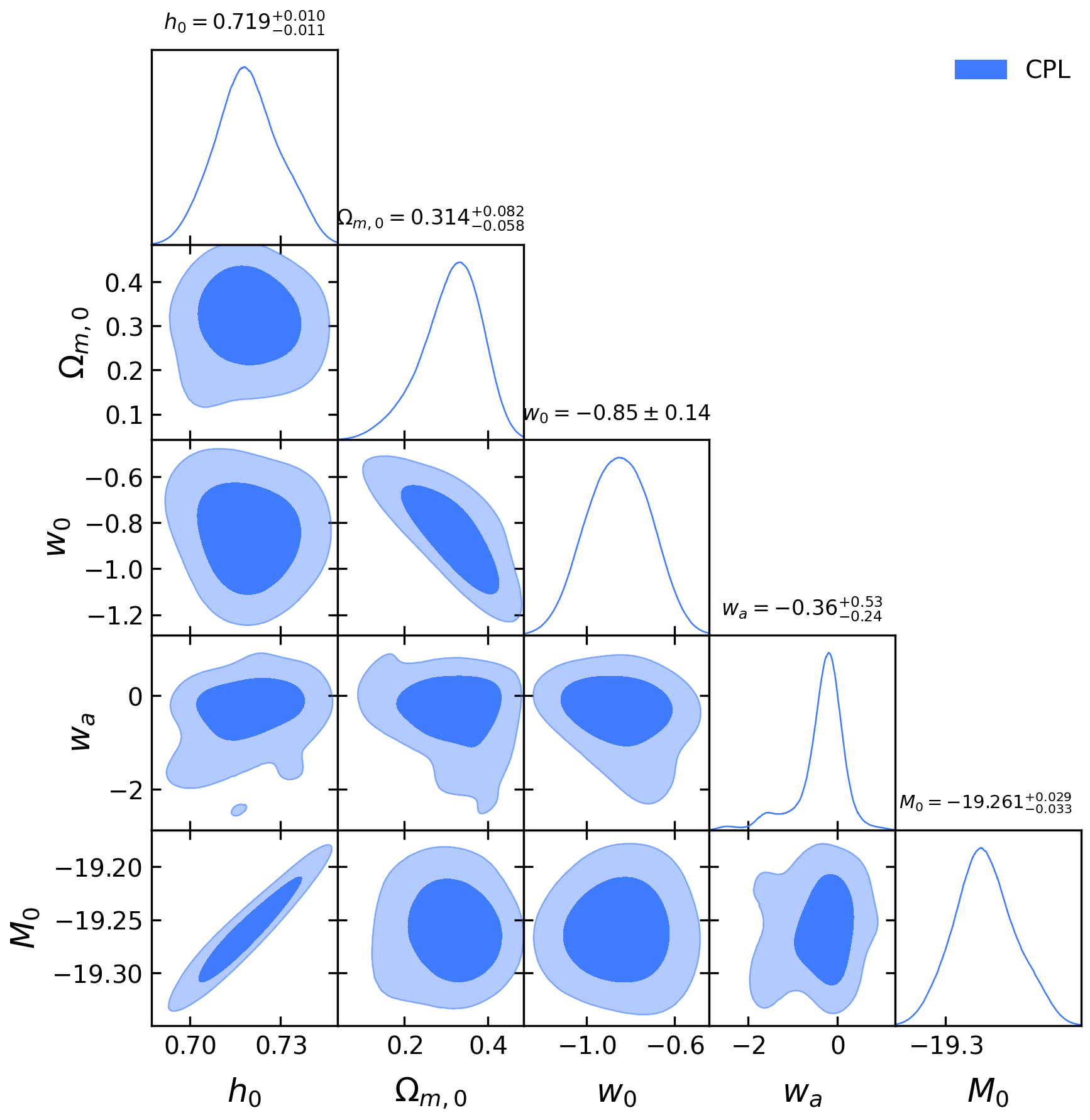}
\includegraphics[width=0.47\linewidth]{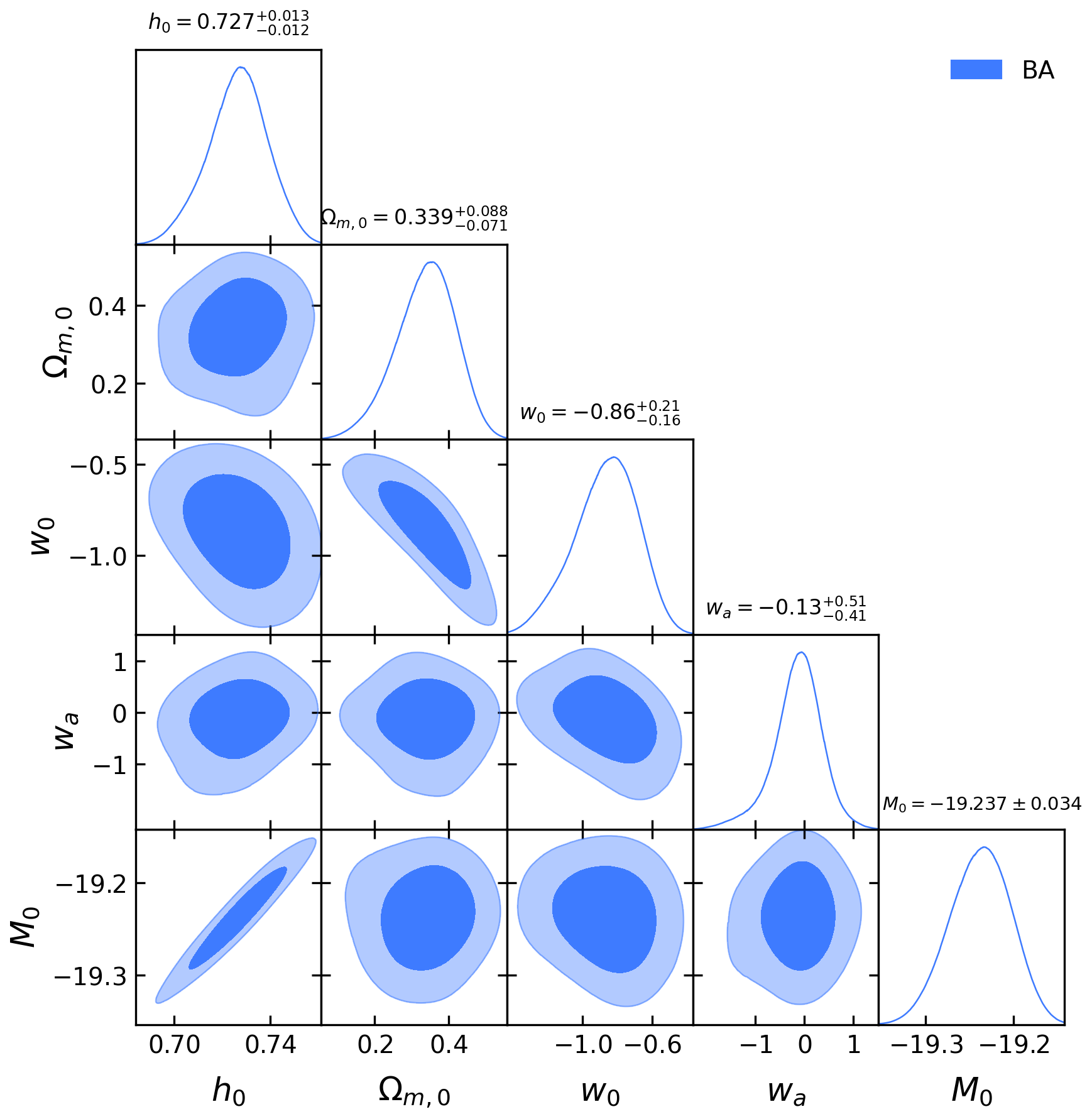}
~
\includegraphics[width=0.47\linewidth]{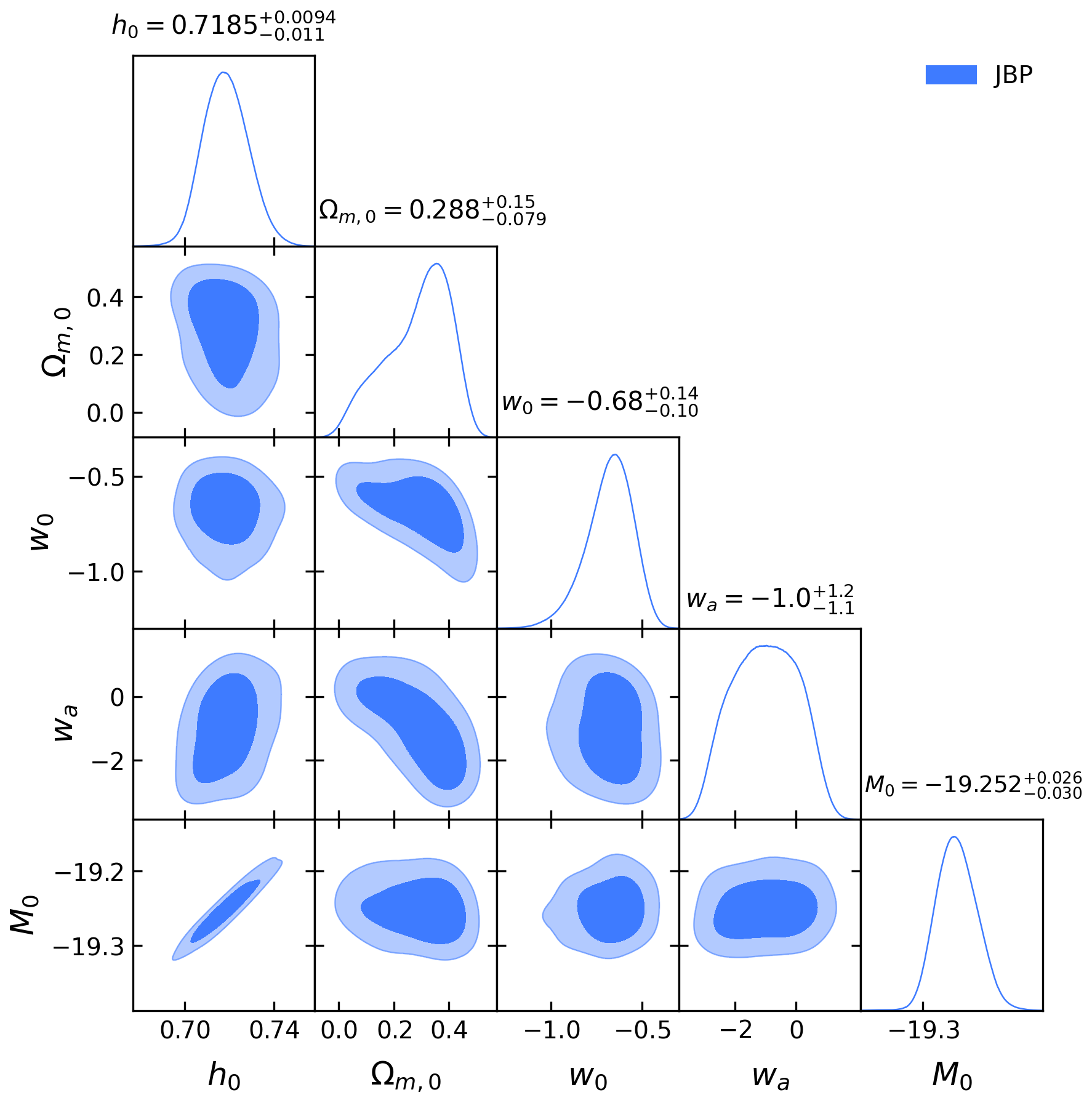}
\includegraphics[width=0.47\linewidth]{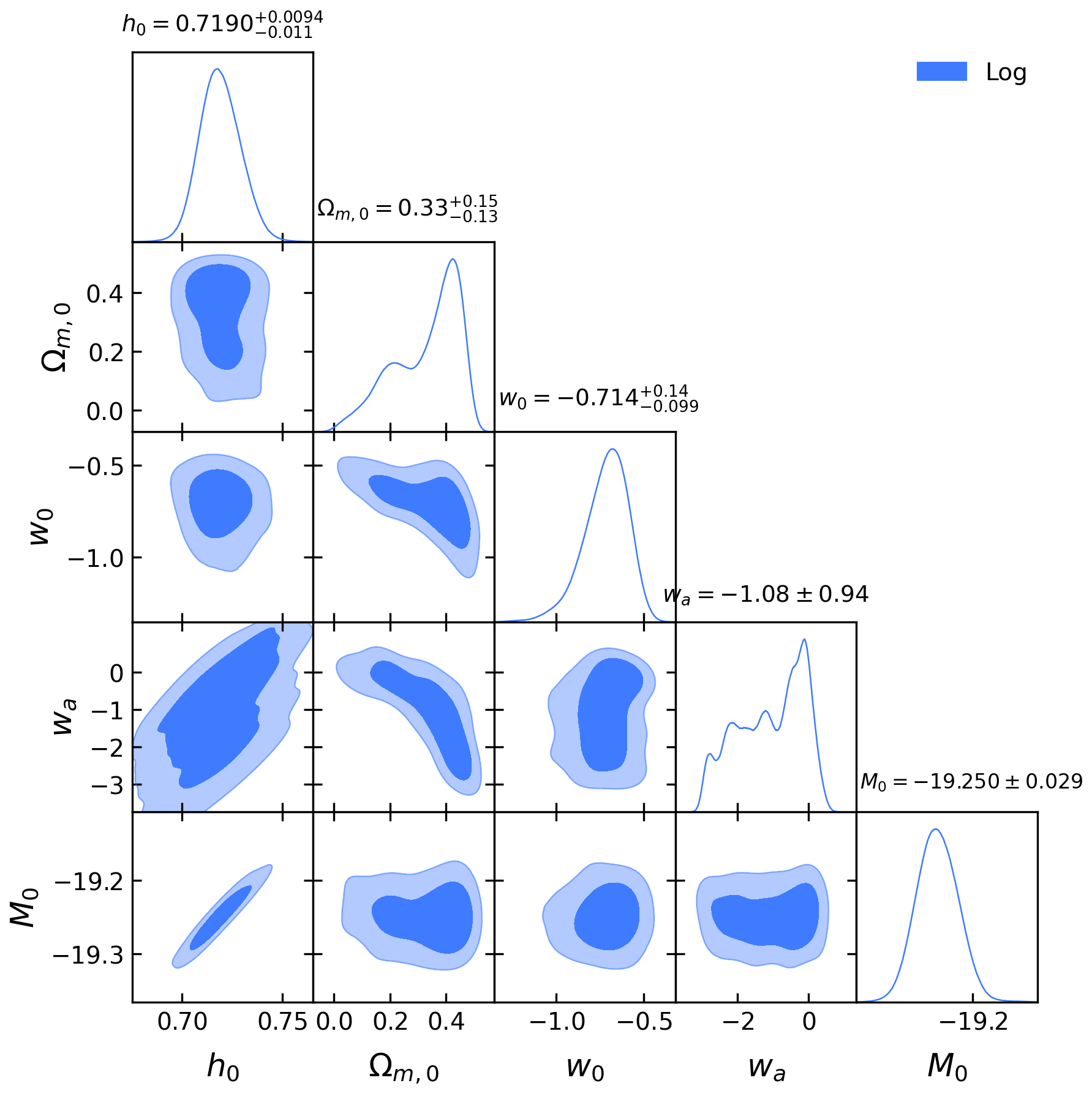}
~
\includegraphics[width=0.47\linewidth]{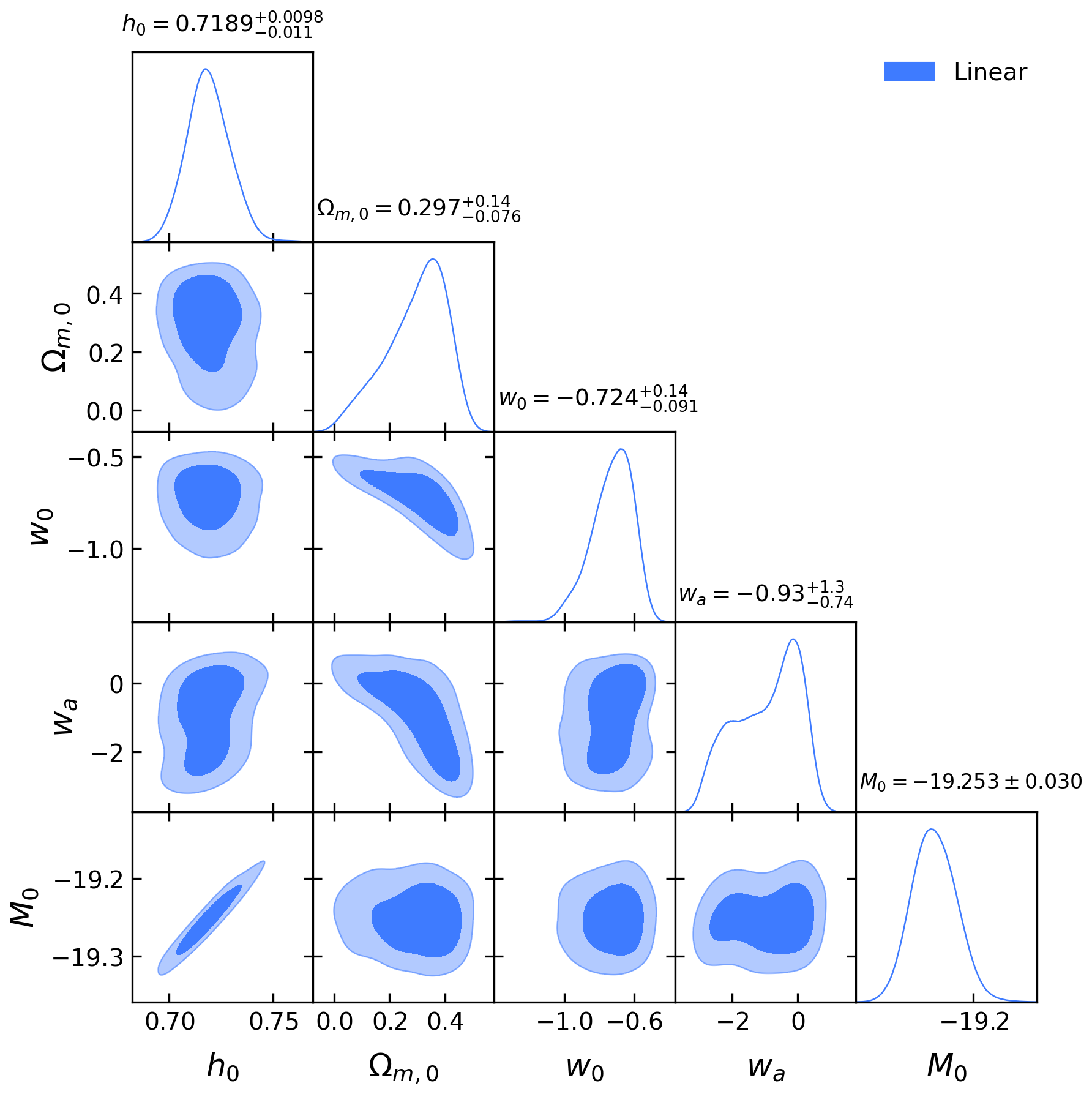}
\caption{2D contour plots for all cosmological parameters considered in all five equation of state parameterizations along with the base $\Lambda$CDM model.} 
\label{fig:all_triangles}
\end{figure}

We perform MCMC sampling for each dark energy EoS parameterization model described earlier, using the Pantheon+ supernova dataset with SH0ES calibrators. The five models - CPL, Barboza–Alcaniz (BA), Jassal–Bagla–Padmanabhan (JBP), Linear-$z$, and Logarithmic-$z$ are compared against the standard $\Lambda$CDM baseline. Each case involves joint inference over the parameter set $\boldsymbol{\Theta} = (h_0, \Omega_{m,0}, w_0, w_a, M_0)$, as described. In Fig.~[\ref{fig:all_triangles}], we present the corner plots showing the marginalised 1D posteriors and 2D contours (with $1\sigma$ and $2\sigma$ levels) for each model. The best-fit values and 68\% ($1\sigma$) credible intervals are summarised in Table \ref{tab:post_corrected}.

An important parameter in every late universe study is the Hubble parameter $H_0$. The values we get for this parameter agrees essentially with the late-universe values as reported in the standard $\Lambda$CDM (we will refer this as base model in this section) constraints by cephied based SNIa analysis done by Pantheon+SH0ES team \cite{Riess:2021shoes}. We get the mean hubble value of $\sim 0.721$ in our case of $\Lambda$CDM, CPL and BA models, while it takes the values $~\sim 0.718$ in remaining three cases of JBP, Linear and Log-linear models. The fractional matter density parameter $\Omega_{m,0}$ is $\sim 0.408$ for the base model is slightly higher than that of other five models $\sim 0.33$, which is understood since we having different equation of state parameterizations and hence this is rather compensation for additional relaxed free parameters $\omega_0, \omega_a$ being introduced in them and also align with the recent Dark Energy Survey (DES) results \cite{DESI_2024_VI}. The SNIa absolute magnitude parameter $M_0$ is smallest for the CPL model, however all of the models provide esssentially similar constraints on it.

The mean $\omega_0$ value in all dynamical–$w$ models (CPL, BA, JBP, Linear‐$z$, Logarithmic-$z$) are greater than the fixed value of `-1' for the base model. However, $\omega_0$ for the first two models (CPL and BA) are $\sim -0.85\pm{0.14}$, it is $\sim -0.72^{0.14}_{-0.09}$ for the last two (Linear and log-linear) ones while $-0.68^{0.14}_{-0.10}$ for the JPB model, suggesting favor for evolving dark energy models \cite{DESI_2024_VI}.

The evolution parameter \( \omega_a \), which governs the redshift dependence of the dark energy equation of state, is unconstrained in the base $\Lambda$CDM model where it is fixed to zero. For the dynamical models, the fitted values vary significantly across parametrizations. In the CPL and BA models, \( \omega_a \) is constrained to be mildly negative, with values \( -0.36^{+0.53}_{-0.24} \) and \( -0.13^{+0.51}_{-0.31} \) respectively. The remaining three models i.e JBP, Linear-$z$, and Logarithmic-$z$ show larger uncertainties and more negative mean values: \( \omega_a \sim -1.0^{+1.2}_{-1.1} \), \( \sim -0.93^{+1.3}_{-0.74} \), and \( \sim -1.08 \pm 0.94 \), respectively. While these are consistent with zero within $1$–$2\sigma$ intervals, they do allow for a more dynamic behavior of dark energy at intermediate redshifts. These wider contours reflect the degeneracies inherent in supernova-only data when attempting to constrain higher-order variations in \( w(z) \), and highlight the complementarity of combining such analyses with other probes, as emphasized in recent DES constraints \cite{DESI_2024_VI}.

Our parameter constraints are essentially in line with a  recent similar work, validating cosmology informed neural networks which only incorporating one such parameterization i.e CPL \cite{2023cinn}. Since one of the major goals of this work was validating the deep-learning techniques in cosmology, it is worth addressing the uncertainties/caveats of the NN modelling as well, which we discuss in the next sub-section.

\begin{figure}
\centering
\includegraphics[width=0.49\textwidth]{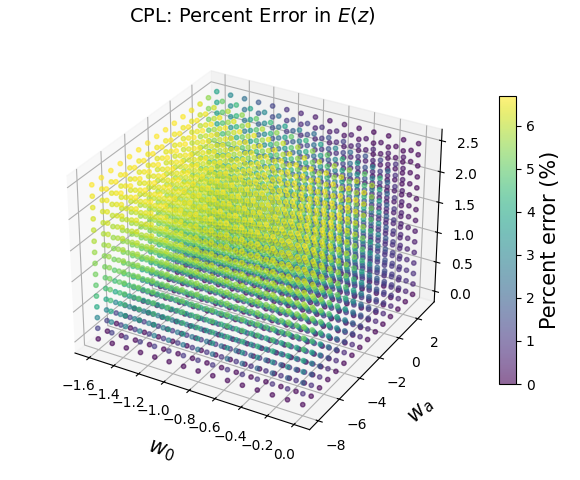}
~
\includegraphics[width=0.49\textwidth]{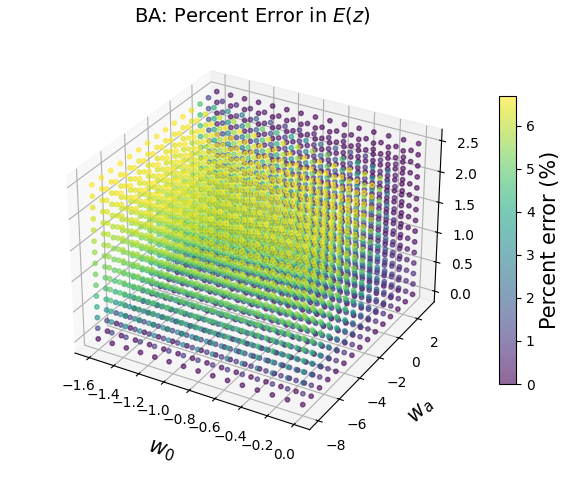}
~
\includegraphics[width=0.49\textwidth]{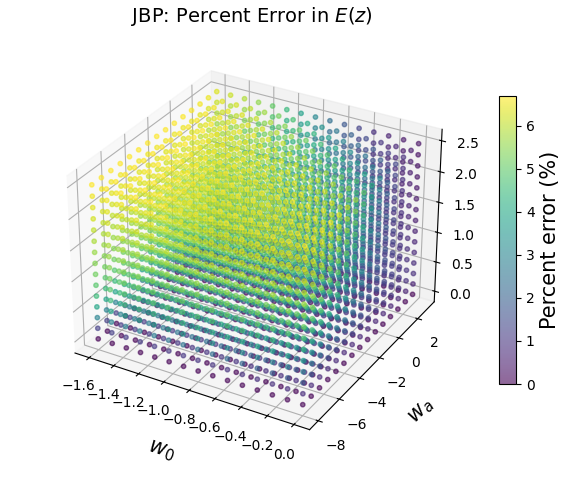}
~
\includegraphics[width=0.49\textwidth]{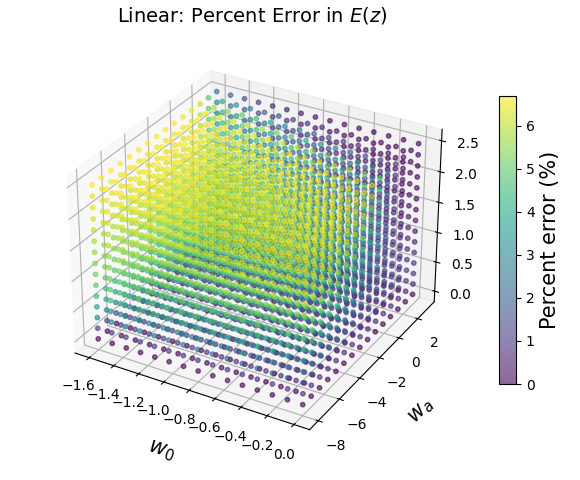}
~
\includegraphics[width=0.49\textwidth]{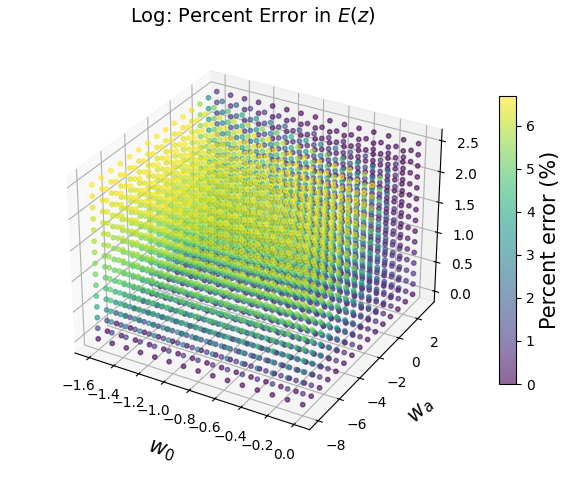}
\caption{Three\mbox{‐}dimensional view of the percent error in the dimensionless Hubble rate $E(z)=H(z)/H_{0}$, obtained by combining the ANN\mbox{‐}based matter term with analytic dark\mbox{‐}energy factors, for five parametrizations (CPL, BA, JBP, Linear, and Logarithmic). The domain is sampled on a uniform $15\times15\times15$ grid in $(w_{0},w_{a},z)$ i.e., $w_{0}\in[-1.6,0.0]$, $w_{a}\in[-8,3]$, and $z\in[0,2.5]$, yielding $15^3=3375$ grid points per panel. At each point $(w_{0},w_{a},z)$, the color encodes the percent error relative to the analytic solution (see color bar); dark blue indicates the smallest errors, whereas yellow/orange marks the largest deviations within the plotted range.}

\label{fig:percent_error}
\end{figure}

\begin{table}
\centering
\begin{tabular}{lcccc}
\hline
Model      &
$H_0$ [km\,s$^{-1}$\,Mpc$^{-1}$] &
$\Omega_{m,0}$ &
$w_0$ &
$w_a$ \\
\hline
$\Lambda$CDM
  & $0.7219 \pm 0.0090$
  & $0.408 \pm 0.021$
  & \multicolumn{1}{c}{(fixed $=-1$)}
  & \multicolumn{1}{c}{(fixed $=0$)}
  \\[0.6ex]
CPL
  & $0.7190^{+0.010}_{-0.011}$
  & $0.314^{+0.082}_{-0.058}$
  & $-0.85 \pm 0.14$
  & $-0.36^{+0.53}_{-0.24}$
  \\[0.6ex]
BA
  & $0.727^{+0.013}_{-0.012}$
  & $0.339^{+0.089}_{-0.088}$
  & $-0.86^{+0.16}_{-0.14}$
  & $-0.13^{+0.51}_{-0.31}$
  \\[0.6ex]
JBP
  & $0.7185^{+0.0094}_{-0.011}$
  & $0.288^{+0.15}_{-0.079}$
  & $-0.68^{+0.14}_{-0.10}$
  & $-1.0^{+1.2}_{-1.1}$
  \\[0.6ex]
Linear-$z$
  & $0.7189^{+0.0098}_{-0.011}$
  & $0.297^{+0.14}_{-0.076}$
  & $-0.724^{+0.14}_{-0.091}$
  & $-0.93^{+1.3}_{-0.74}$
  \\[0.6ex]
Logarithmic-$z$
  & $0.7190^{+0.0094}_{-0.011}$
  & $0.330^{+0.15}_{-0.13}$
  & $-0.714^{+0.14}_{-0.099}$
  & $-1.08 \pm 0.94$
  \\
\hline
\end{tabular}
\caption{Posterior values for 1$\sigma$ credible intervals for all the free parameters for all models considered in our analysis using Pantheon+SH0ES data, whose contour plots are also shown in the Fig.~\ref{fig:all_triangles}}
\label{tab:post_corrected}
\end{table}

\subsection{Accuracy of the ANN Surrogate}
\label{sec:accuracy_results}

\begin{figure}
    \centering
    \includegraphics[width=1\textwidth]{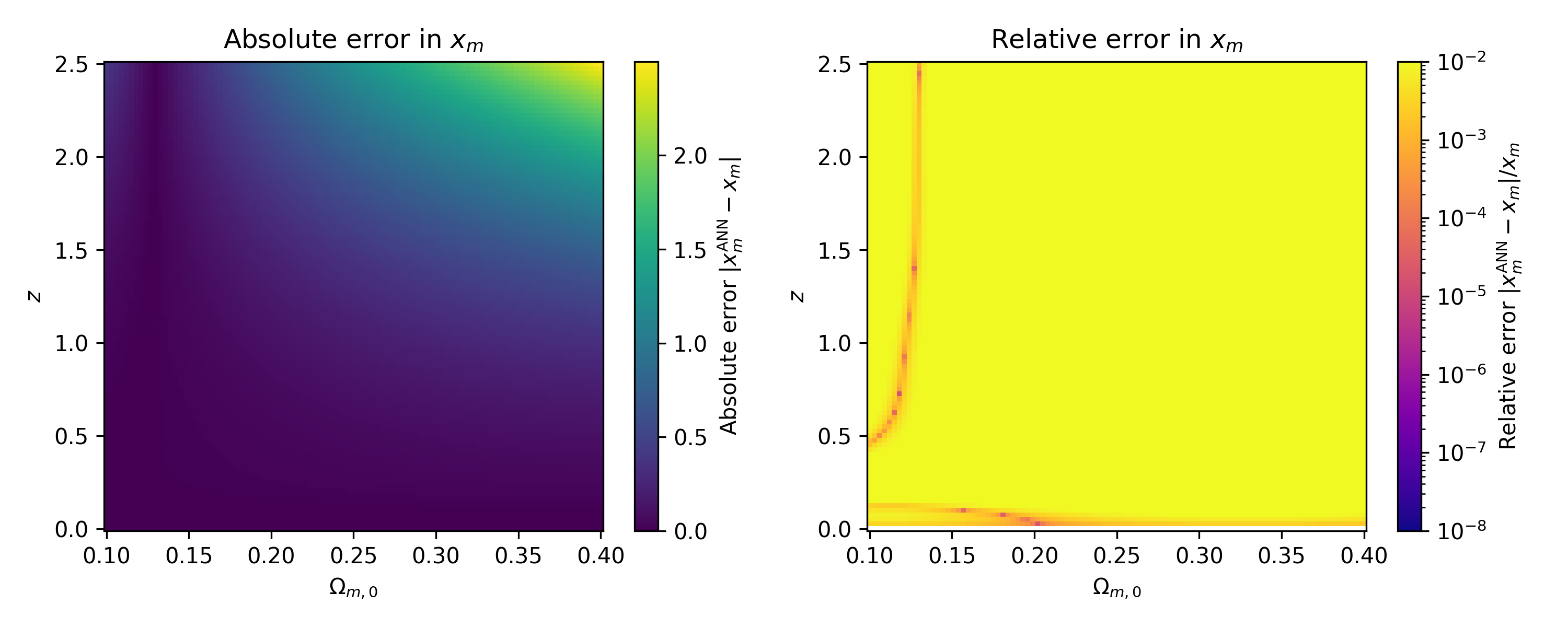}
    \caption{Absolute and relative errors of the surrogate matter term $x_m^{\rm ANN}(z,\Omega_{m,0})$ with respect to the analytic $\Omega_{m,0}(1+z)^3$.  Over the full validation grid $z\in[0,2.5]$ and $\Omega_{m,0}\in[0.1,0.4]$, the relative error remains below $10^{-2}$, and is typically $\mathcal{O}(10^{-5})$.}
    \label{fig:xm_error}
\end{figure}

\begin{figure}[htbp]
  \centering
  \includegraphics[width=1\textwidth]{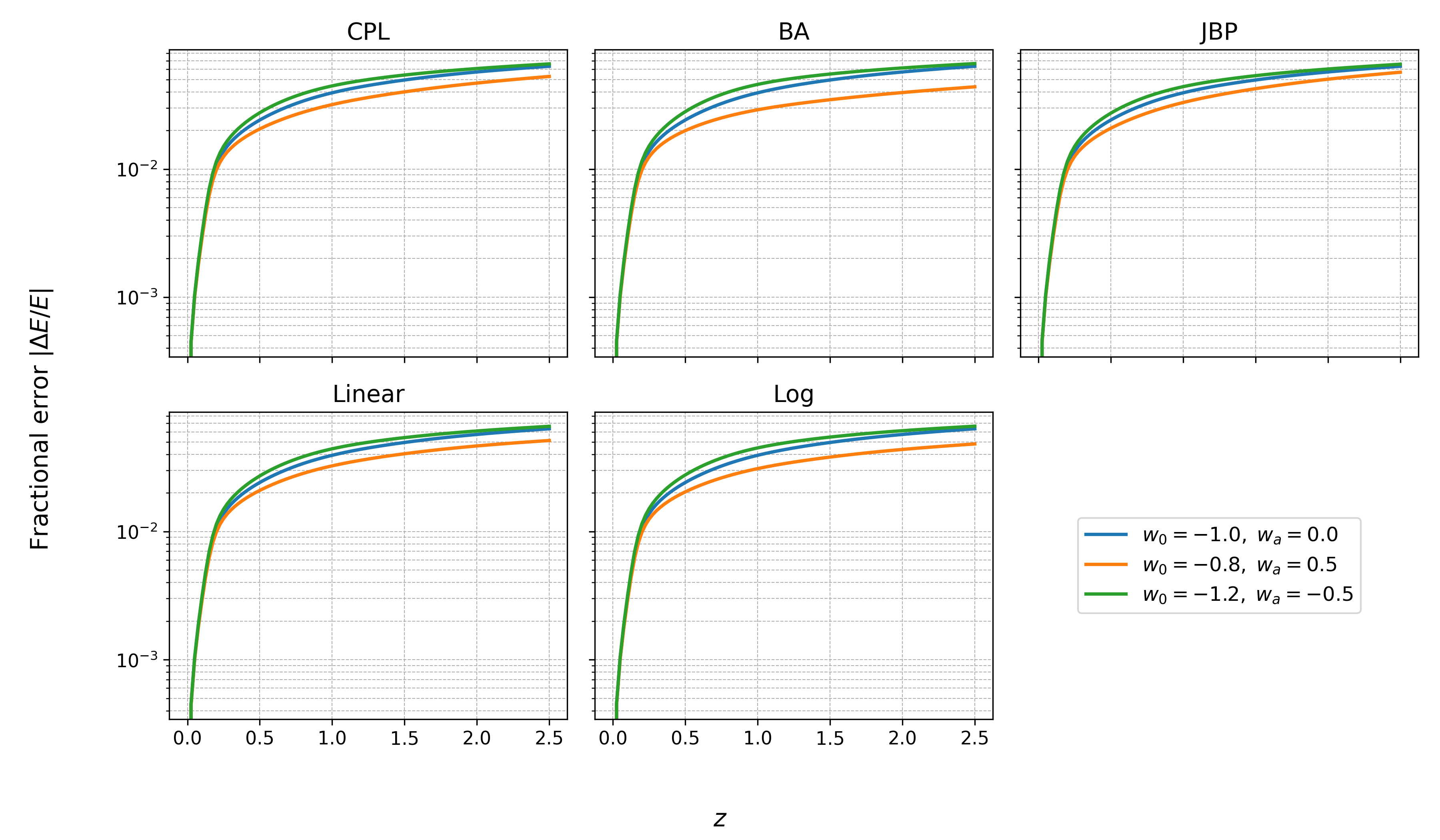}
  \caption{Fractional error in the dimensionless Hubble rate
  $E(z)=H/H_0$ for three representative
  $(w_0,w_a)$ pairs \mbox{$(-1,0)$} (blue),
  \mbox{$(-0.8,0.5)$} (orange), and \mbox{$(-1.2,-0.5)$} (green),
  shown for the CPL, BA, JBP, Linear-$z$, and Logarithmic-$z$ models
  (left to right).  In every case $|\Delta E/E|<10^{-4}$ out to
  $z=2.5$, well below the statistical precision of current SN data.}
\label{fig:Eerr_models}
\end{figure}

\begin{figure}[htbp]
  \centering
  \includegraphics[width=1\linewidth]{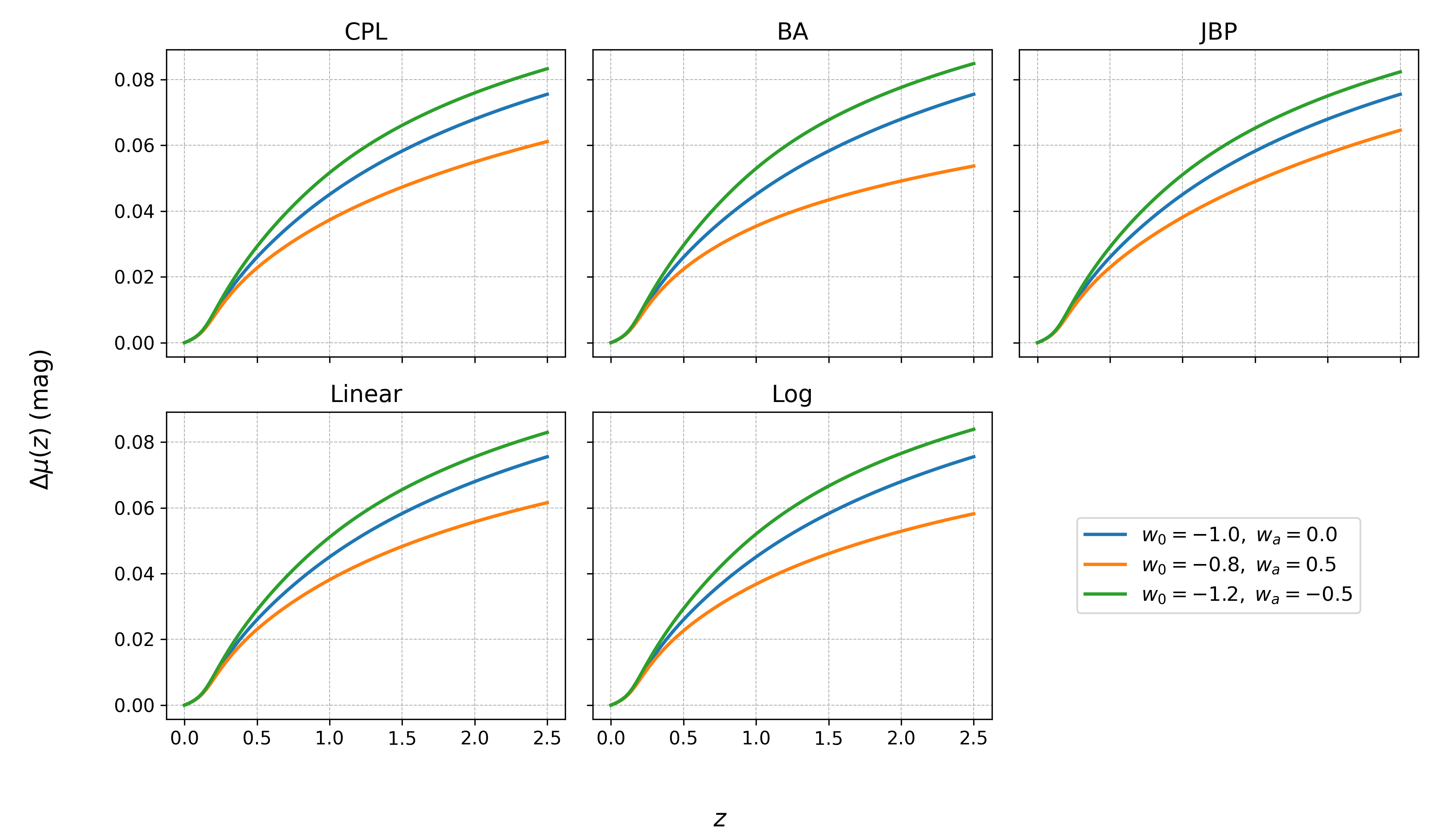}
  \caption{Distance-modulus error
  $\Delta\mu(z)=\mu_{\rm ANN}(z)-\mu_{\rm exact}(z)$ for the same
  parameter triplet as Fig.~[\ref{fig:Eerr_models}].  Even at
  $z\simeq2.5$ the bias stays below $0.1\,\mathrm{mag}$.}
\label{fig:dm_models}
\end{figure}

We now assess the numerical fidelity of our trained ANN surrogates across three key diagnostics relevant for cosmological inference. Fig.~\ref{fig:percent_error} surveys the \emph{full} 3D $(w_{0},w_{a},z)$ grid for each dark energy parametrization and shows that our ANN based surrogate reproduces the analytical function $E(z)$ with percent level accuracy across most of the sampled volume. Darker colors denote smaller errors, while the largest deviations confined to the edges of the previous (e.g., $w_{a}\gtrsim 2$ or $w_{0}\lesssim -1.4$) reach only a few percent, consistent with the $0$–$6\,\%$ range indicated by the color bars. These errors are small compared to our analysis requirements and do not affect downstream parameter inference.

Although $x_m(z;\Omega_{m,0})=\Omega_{m,0}(1+z)^3$ is analytic, we implement it with a compact auxiliary neural network (two hidden layers, $<10^2$ weights) so that the background, matter and dark energy, lies inside a single differentiable PyTorch module for GPU batch evaluation. The network is trained once with an $L_2$ loss,
$\mathcal{L}_{\rm mat}=\langle|x^{\rm ANN}_m-\Omega_{m,0}(1+z)^3|^2\rangle$, over the grid $(z,\Omega_{m,0})\in[0,2.5]\times[0.1,0.4]$.
Figure~\ref{fig:xm_error} shows the resulting residuals: absolute errors increase with $z$ as expected from the $(1+z)^3$ scaling, and relative errors are at most a few $\times10^{-2}$ across the domain (see color bars). These residuals are acceptable for our analysis and are not the dominant source of the total error budget shown in Fig.~\ref{fig:percent_error}.

We propagate the surrogate predictions into the function \( E(z) \) for each $w(z)$ model. Shown in Fig.~[\ref{fig:Eerr_models}] is the fractional error \( \Delta E / E \) for three representative parameter choices. The error remains below \( 10^{-4} \) across all redshifts, which is well within tolerance for current SN-based cosmology. Finally, we compute the induced error on the observable distance modulus \( \mu(z) \). In Fig.~[\ref{fig:dm_models}], it was shown that the surrogate bias \( \Delta \mu(z) = \mu_{\rm ANN}(z) - \mu_{\rm exact}(z) \) remains below \( 0.1\,\mathrm{mag} \) even at \( z \simeq 2.5 \). Since this is smaller than the intrinsic scatter in Pantheon+ data, the surrogate-induced bias is negligible.

\subsection{CINN Efficiency}

In this section, we quantify the computational efficiency of using a Cosmology-Informed Neural Network (CINN) surrogate relative to direct numerical integration within MCMC sampling. We also discuss the conditions under which the surrogate becomes advantageous, as well as its scientific utility. Standard Bayesian inference with MCMC repeatedly evaluates the model prediction
(e.g. the Hubble rate and the luminosity distance) for many parameter draws.
Each draw normally requires numerical integration of the background ODE.
A CINN is trained once to emulate the entire map
\((z,\theta)\mapsto x_{\rm de}(z;\theta)\) (the ``bundle'' of solutions)
and thereafter returns predictions with a single forward pass. This amortises (pays off)
the cost of integration: an upfront training cost replaces a very large number of
per-sample integrations during inference. 

Let $N_{\rm runs}$ be the number of independent Bayesian analyses (different datasets, priors, or samplers) performed for the \emph{same} $w(z)$ model.
The total computational cost for the two approaches is:
\begin{align}
    C_{\rm direct} &= N_{\rm runs} \, T_{\rm MCMC} \\
    C_{\rm surrogate} &= T_{\rm train} + N_{\rm runs} \, T_{\rm sur} \, .
\end{align}
The surrogate becomes cost--effective when
\begin{equation}
    T_{\rm train} < N_{\rm runs} \left( T_{\rm MCMC} - T_{\rm sur} \right) .
    \label{eq:break_even}
\end{equation}
As an example, in our CPL analysis with $(T_{\rm train}, T_{\rm MCMC}, T_{\rm sur}) \simeq (120,\, 40,\, 10)~\mathrm{min}$, Eq.~\eqref{eq:break_even} yields
\begin{equation}
    N_{\rm runs} > \frac{120}{40 - 10} \simeq 4 .
\end{equation}
Thus, after $\gtrsim 4$ independent analyses for the same model, the total wall-clock time with the surrogate is smaller than direct integration for each run. This means that the efficiency of our CPL surrogate model can only be truly judged when the number or independent simulations to be done through standard MCMC for the same model is more that 4. Note that leveraging parallelization can significantly reduce the wall‐clock time required for both training and evaluating the ANNs, making them a more attractive option compared to the earlier scenario. A more detailed and clear explanation of the cost effectivemenss can be found in another recent CINN based work~\cite{2023cinn} (Appendix F).

The scientific significance of the CINNs can similarly be understood by the cost effectivemess in certain scenarios. For $\Lambda$CDM or CPL, the background ODEs are fast to solve and the benefit is modest. However, for more complex models like modified gravity \cite{Shankaranarayanan_2022} and anisotorpic cosmological models \cite{Verma_2025} , the cost per likelihood evaluation can increase by orders of magnitude. In such cases, $T_{\rm MCMC}$ may be hours rather than minutes, and the break-even $N_{\rm runs}$ from Eq.~\eqref{eq:break_even} can drop to $<1$, making the surrogate advantageous even for a single run. The CINN surrogate may offer possible advantages beyond to wall clock savings too. The trained network can be shared as a standalone background evolution module for
CLASS, CAMB, or other Boltzmann solvers (for a fixed model, say, $\Lambda$CDM). Other researchers can plug it into their pipelines without
re implementing the ODE solver.

However, the cosmological precision for the surrogate outputs would most likely be unreliable for scientific purposes. For simple background only models like $\Lambda$CDM and CPL with a single dataset, the surrogate does not
provide a net time saving over direct integration. Its main value lies in repeated analyses for the same model, complex dynamical models where $T_{\rm MCMC} \gg T_{\rm train}$ and enabling inference workflows that demand differentiable, vectorized model evaluations. Also, one can use different data set, for example Dark Energy Survey (DES) SNe dataset~\cite{descollaboration2025darkenergysurveycosmology}, for a given trained model surrogate to infer the posteriors. In that case, one should be aware that if the new dataset has $z$ outside your training range, you either need to retrain or extend the training grid.

All of these comes with the caveat of being unreliable while doing a precise cosmological inference but this issue can be improved upon with either the technological advancement in PINN techniques in general or with availability of better quality datasets suitable for the cosmological model in question. In addition, the surrogate
provides differentiability and GPU-batched evaluation, enabling gradient-based
samplers and large-scale parameter surveys that further increase the relative
benefit of the surrogate approach.

\section{Conclusions}
\label{conclusion}

In this work, we have implemented and validated a deep-learning-based framework for background cosmology inference using the Pantheon+ supernova dataset with SH0ES calibration. We explored six cosmological models in total: the standard $\Lambda$CDM model and five evolving dark energy scenarios based on widely studied equation-of-state parametrizations - CPL, Barboza–Alcaniz (BA), Jassal–Bagla–Padmanabhan (JBP), Linear-$z$, and Logarithmic-$z$. All of these models were constrained via MCMC sampling using a physics-informed neural network (PINN) surrogate as a fast and differentiable replacement for quadrature-based numerical solvers. The surrogate was designed to reproduce the Hubble function $E(z)$ and the observable distance modulus $\mu(z)$ to extremely high accuracy over the relevant redshift and parameter space. Our results show that the surrogate model achieves sub-per-mille precision across the $(w_0, w_a, z)$ space for all parametrizations, and that the induced errors in $\mu(z)$ remain well below the intrinsic dispersion of the data. Consequently, the neural network approximation does not introduce any significant bias in parameter estimation and can be safely used in likelihood-based cosmological inference.

From the parameter constraints obtained, we find that all five dynamical dark energy models yield results consistent with the $\Lambda$CDM baseline at the $95\%$ confidence level. The $w_0$ and $w_a$ parameters in these models prefer higher than the base $\Lambda$CDM model, for with there is no redshift evolution and $\omega$ is simply fixed to `-1' , having mild deviations within themselves in the best-fit values for the JBP and Logarithmic-$z$ models. However, the posterior volumes are large enough that these are not statistically significant. In particular, our CPL constraints $(w_0 = -0.85 \pm 0.14,\, w_a = -0.36^{+0.53}_{-0.24})$ are in agreement with recent results of higher values of $\omega_0$ from DES-Y3 analysis when combined with external datasets.

Importantly, the inferred value of the Hubble constant $H_0$ remains in agreement with late-universe measurements from SH0ES in all models. The $\Lambda$CDM baseline gives $H_0 \simeq 72.2\,{\rm km\,s^{-1}\,Mpc^{-1}}$, while the dynamical models return values in the narrow range $H_0 \sim 71.8$–$72.7$, all within the calibration uncertainty. The matter density $\Omega_{m,0}$ is slightly reduced in models with evolving $w(z)$, consistent with the additional freedom allowed by a dynamic dark energy sector.

These results confirm that, at current precision, the Pantheon+ Type Ia supernovae support a $w$CDM model with $w>-1$, although dynamical dark energy cannot be ascertained due to weak constraints on the $w_a$ parameter from SNe Ia data. Crucially, our PINN‐based surrogate framework reduces the computational cost (few hours) of exploring multiple parameterizations by orders of magnitude, making it straightforward to extend this analysis to include additional probes (e.g., BAO, cosmic chronometers, cosmic microwave background) or to test more complex dark energy and modified gravity scenarios.

In summary, while our CINN surrogates do not provide a computational advantage for single run analyses of simple background cosmologies such as $\Lambda$CDM or CPL with the Pantheon+SH0ES data, they become competitive when the same model is analysed repeatedly with different datasets, priors, or sampling configurations. For our CPL case, the break--even point occurs at $\gtrsim 4$ independent runs, beyond which the total wall--clock time is smaller than that of direct numerical integration within MCMC. The efficiency gain is even more pronounced for complex dynamical models (e.g.\ modified gravity, anisotropic cosmologies) where the per likelihood cost can be orders of magnitude higher, making the surrogate advantageous even for a single inference run. Beyond time savings, the trained CINNs `might' be distributed as reusable, differentiable modules for incorporation into Boltzmann solvers like \textsc{CLASS} or \textsc{CAMB}, enabling GPU accelerated, vectorized evaluations for gradient based samplers and large scale parameter scans which is a future objective. These benefits are accompanied by caveats - precision for cosmological inference must be carefully validated, particularly if applying the surrogate to datasets with redshift coverage outside the training range, in which case retraining or grid extension is required. Nonetheless, the approach provides a flexible, shareable, and computationally scalable tool that can facilitate future analyses, especially in regimes where conventional ODE based likelihood evaluations are prohibitively expensive.

\section*{Acknowledgements}
AV acknowledges the financial support received through the Research Fellowship awarded by the Council of Scientific \& Industrial Research (CSIR), India, during the course of this project. SS acknowledges the financial support from Washington University in St. Louis during the course of this project. DFM thanks the Research Council of Norway for their support and the resources provided by UNINETT Sigma2 -- the National Infrastructure for High-Performance Computing and Data Storage in Norway. We acknowledge National Supercomputing Mission (NSM) for providing computing resources of `PARAM
Shivay' at Indian Institute of Technology (BHU), Varanasi, which is implemented by C-DAC and supported by the Ministry of Electronics and Information Technology (MeitY) and Department of Science and Technology (DST), Government of India.
Software acknowledgments:
\texttt{Neural Ordinary Differntial Equations}\footnote{\url{https://github.com/NeuroDiffGym/neurodiffeq}}~\cite{2018arXiv180607366C},
\texttt{Cobaya}\footnote{\url{https://cobaya.readthedocs.io/}}~\cite{torrado2021},
\texttt{emcee}\footnote{\url{https://ui.adsabs.harvard.edu/abs/2013ascl.soft03002F}}~\cite{2013PASP..125..306F},
\texttt{GetDist}\footnote{\url{https://cobaya.readthedocs.io/en/latest/}}~\cite{lewis:2019xzd},
\texttt{SciPy}\footnote{\url{https://scipy.org/}}~\cite{scipy2020},
\texttt{NumPy}\footnote{\url{https://numpy.org/}}~\cite{numpy2020},
\texttt{Astropy}\footnote{\url{http://www.astropy.org}}~\cite{astropy2013,astropy2018,astropy2022},
and \texttt{Matplotlib}\footnote{\url{https://matplotlib.org/}}~\cite{hunter:2007}.

\bibliographystyle{unsrt}
\bibliography{ref_new}

\begin{thebibliography}{10}

\bibitem{Goodfellow-et-al-2016}
Ian Goodfellow, Yoshua Bengio, and Aaron Courville.
\newblock {\em Deep Learning}.
\newblock MIT Press, 2016.
\newblock \url{http://www.deeplearningbook.org}.

\bibitem{yegnanarayana2009artificial}
Bayya Yegnanarayana.
\newblock {\em Artificial neural networks}.
\newblock PHI Learning Pvt. Ltd., 2009.

\bibitem{Uhrig_ANN_1995}
R.E. Uhrig.
\newblock Introduction to artificial neural networks.
\newblock In {\em Proceedings of IECON '95 - 21st Annual Conference on IEEE Industrial Electronics}, volume~1, pages 33--37 vol.1, 1995.

\bibitem{krenker2011introduction}
Andrej Krenker, Janez Be{\v{s}}ter, and Andrej Kos.
\newblock Introduction to the artificial neural networks.
\newblock {\em Artificial Neural Networks: Methodological Advances and Biomedical Applications. InTech}, pages 1--18, 2011.

\bibitem{raissi2019physics}
M.~Raissi, P.~Perdikaris, and G.E. Karniadakis.
\newblock Physics-informed neural networks: A deep learning framework for solving forward and inverse problems involving nonlinear partial differential equations.
\newblock {\em Journal of Computational Physics}, 378:686--707, 2019.

\bibitem{2025PhLB..86839690B}
M.~P. {Bento}, H.~B. {C{\^a}mara}, and J.~F. {Seabra}.
\newblock {Unraveling particle dark matter with Physics-Informed Neural Networks}.
\newblock {\em Physics Letters B}, 868:139690, September 2025.

\bibitem{2015JCAP...09..045V}
S.~D.~P. {Vitenti} and M.~{Penna-Lima}.
\newblock {A general reconstruction of the recent expansion history of the universe}.
\newblock {\em \jcap}, 2015(9):045--045, September 2015.

\bibitem{2022MNRAS.512L..44S}
A.~{Spurio Mancini} and A.~{Pourtsidou}.
\newblock {KiDS-1000 cosmology: machine learning - accelerated constraints on interacting dark energy with COSMOPOWER}.
\newblock {\em \mnras}, 512(1):L44--L48, May 2022.

\bibitem{2020arXiv200614372F}
Cedric {Flamant}, Pavlos {Protopapas}, and David {Sondak}.
\newblock {Solving Differential Equations Using Neural Network Solution Bundles}.
\newblock {\em arXiv e-prints}, page arXiv:2006.14372, June 2020.

\bibitem{2023cinn}
Augusto~T. {Chantada}, Susana~J. {Landau}, Pavlos {Protopapas}, Claudia~G. {Sc{\'o}ccola}, and Cecilia {Garraffo}.
\newblock {Cosmology-informed neural networks to solve the background dynamics of the Universe}.
\newblock {\em \prd}, 107(6):063523, March 2023.

\bibitem{Perlmutter:1999}
S.~Perlmutter et~al.
\newblock {Measurements of {\ensuremath{\Omega}} and {\ensuremath{\Lambda}} from 42 High-Redshift Supernovae}.
\newblock {\em Astrophys. J.}, 517:565--586, 1999.

\bibitem{Riess:2021shoes}
Adam~G. Riess et~al.
\newblock {A Comprehensive Measurement of the Local Value of the Hubble Constant with 1 km s$^{-1}$ Mpc$^{-1}$ Uncertainty from the Hubble Space Telescope and the SH0ES Team}.
\newblock {\em Astrophys. J. Lett.}, 934(1):L7, 2022.

\bibitem{2022ppluscosmo}
Dillon Brout et~al.
\newblock {The Pantheon+ Analysis: Cosmological Constraints}.
\newblock {\em Astrophys. J.}, 938(2):110, 2022.

\bibitem{CarrollPressTurner1992}
Sean~M. {Carroll}, William~H. {Press}, and Edwin~L. {Turner}.
\newblock {The cosmological constant.}
\newblock {\em \araa}, 30:499--542, January 1992.

\bibitem{2000astro.ph..5265W}
Steven {Weinberg}.
\newblock {The Cosmological Constant Problems (Talk given at Dark Matter 2000, February, 2000)}.
\newblock {\em arXiv e-prints}, pages astro--ph/0005265, May 2000.

\bibitem{Carroll2001LRR}
Sean~M. {Carroll}.
\newblock {The Cosmological Constant}.
\newblock {\em Living Reviews in Relativity}, 4(1):1, December 2001.

\bibitem{PeeblesRatra2003RvMP}
P.~J. {Peebles} and Bharat {Ratra}.
\newblock {The cosmological constant and dark energy}.
\newblock {\em Reviews of Modern Physics}, 75(2):559--606, April 2003.

\bibitem{2013arXiv1309.4133B}
C.~P. {Burgess}.
\newblock {The Cosmological Constant Problem: Why it's hard to get Dark Energy from Micro-physics}.
\newblock {\em arXiv e-prints}, page arXiv:1309.4133, September 2013.

\bibitem{2014arXiv1407.2086M}
J.~W. {Moffat}.
\newblock {Quantum Gravity and the Cosmological Constant Problem}.
\newblock {\em arXiv e-prints}, page arXiv:1407.2086, July 2014.

\bibitem{Planck:2018cosmopar}
N.~Aghanim et~al.
\newblock {Planck 2018 results. VI. Cosmological parameters}.
\newblock {\em Astron. Astrophys.}, 641:A6, 2020.
\newblock [Erratum: Astron.Astrophys. 652, C4 (2021)].

\bibitem{Efstathiou:2020}
G.~{Efstathiou}.
\newblock {A Lockdown Perspective on the Hubble Tension (with comments from the SH0ES team)}.
\newblock {\em arXiv e-prints}, page arXiv:2007.10716, July 2020.

\bibitem{freedman2021}
Wendy~L. {Freedman}.
\newblock {Measurements of the Hubble Constant: Tensions in Perspective}.
\newblock {\em \apj}, 919(1):16, September 2021.

\bibitem{valentino:2021}
Eleonora {Di Valentino}, Olga {Mena}, Supriya {Pan}, Luca {Visinelli}, Weiqiang {Yang}, Alessandro {Melchiorri}, David~F. {Mota}, Adam~G. {Riess}, and Joseph {Silk}.
\newblock {In the realm of the Hubble tension-a review of solutions}.
\newblock {\em Classical and Quantum Gravity}, 38(15):153001, July 2021.

\bibitem{2023Kamionkowski}
Marc {Kamionkowski} and Adam~G. {Riess}.
\newblock {The Hubble Tension and Early Dark Energy}.
\newblock {\em Annual Review of Nuclear and Particle Science}, 73:153--180, September 2023.

\bibitem{2001IJMPD..10..213C}
Michel {Chevallier} and David {Polarski}.
\newblock {Accelerating Universes with Scaling Dark Matter}.
\newblock {\em International Journal of Modern Physics D}, 10(2):213--223, January 2001.

\bibitem{2003PhRvL..90i1301L}
Eric~V. {Linder}.
\newblock {Exploring the Expansion History of the Universe}.
\newblock {\em \prl}, 90(9):091301, March 2003.

\bibitem{2004PhRvD..70h3007S}
Rom{\'a}n {Scoccimarro}.
\newblock {Redshift-space distortions, pairwise velocities, and nonlinearities}.
\newblock {\em \prd}, 70(8):083007, October 2004.

\bibitem{linden2008test}
Sebastian Linden and Jean-Marc Virey.
\newblock Test of the chevallier-polarski-linder parametrization for rapid dark energy equation<? format?> of state transitions.
\newblock {\em Physical Review D—Particles, Fields, Gravitation, and Cosmology}, 78(2):023526, 2008.

\bibitem{raveri2020reconstructing}
Marco Raveri.
\newblock Reconstructing gravity on cosmological scales.
\newblock {\em Physical Review D}, 101(8):083524, 2020.

\bibitem{alam2021completed}
Shadab Alam et~al.
\newblock {Completed SDSS-IV extended Baryon Oscillation Spectroscopic Survey: Cosmological implications from two decades of spectroscopic surveys at the Apache Point Observatory}.
\newblock {\em Phys. Rev. D}, 103(8):083533, 2021.

\bibitem{lodha2025extended}
K.~Lodha et~al.
\newblock {Extended Dark Energy analysis using DESI DR2 BAO measurements}.
\newblock 3 2025.

\bibitem{2008PhLB..666..415B}
E.~M. {Barboza} and J.~S. {Alcaniz}.
\newblock {A parametric model for dark energy}.
\newblock {\em Physics Letters B}, 666(5):415--419, September 2008.

\bibitem{2003PhRvD..67f3504B}
J.~S. {Bagla}, H.~K. {Jassal}, and T.~{Padmanabhan}.
\newblock {Cosmology with tachyon field as dark energy}.
\newblock {\em \prd}, 67(6):063504, March 2003.

\bibitem{2005MNRAS.356L..11J}
H.~K. {Jassal}, J.~S. {Bagla}, and T.~{Padmanabhan}.
\newblock {WMAP constraints on low redshift evolution of dark energy}.
\newblock {\em \mnras}, 356(1):L11--L16, January 2005.

\bibitem{2001PhRvD..64l3527H}
Dragan {Huterer} and Michael~S. {Turner}.
\newblock {Probing dark energy: Methods and strategies}.
\newblock {\em \prd}, 64(12):123527, December 2001.

\bibitem{2002PhRvD..65j3512W}
Jochen {Weller} and Andreas {Albrecht}.
\newblock {Future supernovae observations as a probe of dark energy}.
\newblock {\em \prd}, 65(10):103512, May 2002.

\bibitem{2018PhRvD..97l3525D}
Zahra {Davari}, Mohammad {Malekjani}, and Michal {Artymowski}.
\newblock {New parametrization for unified dark matter and dark energy}.
\newblock {\em \prd}, 97(12):123525, June 2018.

\bibitem{Feng_2011}
Lei Feng and Tan Lu.
\newblock A new equation of state for dark energy model.
\newblock {\em Journal of Cosmology and Astroparticle Physics}, 2011(11):034–034, November 2011.

\bibitem{2011PhLB..699..233M}
Jing-Zhe {Ma} and Xin {Zhang}.
\newblock {Probing the dynamics of dark energy with novel parametrizations}.
\newblock {\em Physics Letters B}, 699(4):233--238, May 2011.

\bibitem{2017JCAP...06..012T}
Ashutosh {Tripathi}, Archana {Sangwan}, and H.~K. {Jassal}.
\newblock {Dark energy equation of state parameter and its evolution at low redshift}.
\newblock {\em \jcap}, 2017(6):012, June 2017.

\bibitem{2024arXiv240514099C}
Chuqi {Chen}, Yahong {Yang}, Yang {Xiang}, and Wenrui {Hao}.
\newblock {Automatic Differentiation is Essential in Training Neural Networks for Solving Differential Equations}.
\newblock {\em arXiv e-prints}, page arXiv:2405.14099, May 2024.

\bibitem{2012arXiv1205.2653C}
Corinna {Cortes}, Mehryar {Mohri}, and Afshin {Rostamizadeh}.
\newblock {L2 Regularization for Learning Kernels}.
\newblock {\em arXiv e-prints}, page arXiv:1205.2653, May 2012.

\bibitem{scolnic2022pantheon+}
Dan Scolnic, Dillon Brout, Anthony Carr, Adam~G Riess, Tamara~M Davis, Arianna Dwomoh, David~O Jones, Noor Ali, Pranav Charvu, Rebecca Chen, et~al.
\newblock The pantheon+ analysis: the full data set and light-curve release.
\newblock {\em The Astrophysical Journal}, 938(2):113, 2022.

\bibitem{2022ApJ...938..113S}
Dan Scolnic et~al.
\newblock {The Pantheon+ Analysis: The Full Data Set and Light-curve Release}.
\newblock {\em \apj}, 938(2):113, October 2022.

\bibitem{coughlin2018testing}
Michael~W. {Coughlin}, Susana {Deustua}, Augustin {Guyonnet}, Nicholas {Mondrik}, Joseph~P. {Rice}, Christopher~W. {Stubbs}, and John~T. {Woodward}.
\newblock {Testing of the LSST's photometric calibration strategy at the CTIO 0.9 meter telescope}.
\newblock In {\em Observatory Operations: Strategies, Processes, and Systems VII}, volume 10704 of {\em Society of Photo-Optical Instrumentation Engineers (SPIE) Conference Series}, page 1070420, July 2018.

\bibitem{2018A&A...619A.180M}
J.~{Ma{\'\i}z Apell{\'a}niz} and M.~{Weiler}.
\newblock {Reanalysis of the Gaia Data Release 2 photometric sensitivity curves using HST/STIS spectrophotometry}.
\newblock {\em \aap}, 619:A180, November 2018.

\bibitem{2013PASP..125..306F}
Daniel {Foreman-Mackey}, David~W. {Hogg}, Dustin {Lang}, and Jonathan {Goodman}.
\newblock {emcee: The MCMC Hammer}.
\newblock {\em \pasp}, 125(925):306, March 2013.

\bibitem{DESI_2024_VI}
A.~G. Adame et~al.
\newblock {DESI 2024 VI: cosmological constraints from the measurements of baryon acoustic oscillations}.
\newblock {\em JCAP}, 02:021, 2025.

\bibitem{Shankaranarayanan_2022}
S.~Shankaranarayanan and Joseph~P. Johnson.
\newblock Modified theories of gravity: Why, how and what?
\newblock {\em General Relativity and Gravitation}, 54(5), May 2022.

\bibitem{Verma_2025}
Anshul Verma, Pavan~K. Aluri, and David~F. Mota.
\newblock Anisotropic universe with anisotropic dark energy.
\newblock {\em Physical Review D}, 111(8), April 2025.

\bibitem{descollaboration2025darkenergysurveycosmology}
T.~M.~C. Abbott et~al.
\newblock {The Dark Energy Survey: Cosmology Results with {\ensuremath{\sim}}1500 New High-redshift Type Ia Supernovae Using the Full 5 yr Data Set}.
\newblock {\em Astrophys. J. Lett.}, 973(1):L14, 2024.

\bibitem{2018arXiv180607366C}
Ricky T.~Q. {Chen}, Yulia {Rubanova}, Jesse {Bettencourt}, and David {Duvenaud}.
\newblock {Neural Ordinary Differential Equations}.
\newblock {\em arXiv e-prints}, page arXiv:1806.07366, June 2018.

\bibitem{torrado2021}
Jes{\'u}s {Torrado} and Antony {Lewis}.
\newblock {Cobaya: code for Bayesian analysis of hierarchical physical models}.
\newblock {\em \jcap}, 2021(5):057, May 2021.

\bibitem{lewis:2019xzd}
Antony Lewis.
\newblock {GetDist: a Python package for analysing Monte Carlo samples}.
\newblock 2019.

\bibitem{scipy2020}
Pauli Virtanen et~al.
\newblock {{SciPy} 1.0: Fundamental Algorithms for Scientific Computing in Python}.
\newblock {\em Nature Methods}, 17:261--272, 2020.

\bibitem{numpy2020}
Charles~R. Harris et~al.
\newblock Array programming with {NumPy}.
\newblock {\em Nature}, 585(7825):357--362, September 2020.

\bibitem{astropy2013}
T.~P. {Robitaille} et~al.
\newblock {Astropy: A community Python package for astronomy}.
\newblock {\em \aap}, 558:A33, October 2013.

\bibitem{astropy2018}
A.~M. {Price-Whelan} et~al.
\newblock {The Astropy Project: Building an Open-science Project and Status of the v2.0 Core Package}.
\newblock {\em \aj}, 156(3):123, September 2018.

\bibitem{astropy2022}
Adrian~M. {Price-Whelan} et~al.
\newblock {The Astropy Project: Sustaining and Growing a Community-oriented Open-source Project and the Latest Major Release (v5.0) of the Core Package}.
\newblock {\em apj}, 935(2):167, August 2022.

\bibitem{hunter:2007}
J.~D. Hunter.
\newblock Matplotlib: A 2d graphics environment.
\newblock {\em Computing in Science \& Engineering}, 9(3):90--95, 2007.

\end{thebibliography}

\end{document}